\documentclass{article}

\usepackage{arxiv}

\usepackage[utf8]{inputenc} 
\usepackage[T1]{fontenc}    
\usepackage{hyperref}       
\usepackage{url}            
\usepackage{booktabs}       
\usepackage{amsfonts}       
\usepackage{nicefrac}       
\usepackage{microtype}      
\usepackage{graphicx}
\usepackage{textgreek}
\usepackage{amsmath}
\usepackage{multirow}
\usepackage{graphicx}
\usepackage{subcaption}
\usepackage{float}
\usepackage{authblk}
\DeclareUnicodeCharacter{2212}{-}

\usepackage[resetlabels,labeled]{multibib}
\newcites{S}{References}

\makeatletter
\g@addto@macro{\UrlBreaks}{\UrlOrds}
\makeatother

\DeclareUnicodeCharacter{2215}{/}

\title{Memristive Memory Enhancement by Device Miniaturization for Neuromorphic Computing}

\author[1,2,*]{ A. S. Goossens }
\author[1,2]{ M. Ahmadi }
\author[1,2]{ D. Gupta }
\author[1,2]{ I. Bhaduri }
\author[1,2]{ B. J. Kooi }
\author[1,2.*]{ T. Banerjee }
\affil[1]{Groningen Cognitive Systems and Materials Center, University of Groningen, Groningen, The Netherlands}
\affil[2]{Zernike Institute for Advanced Materials, University of Groningen, Groningen, The Netherlands}
\affil[*]{\texttt{\{a.s.goossens,t.banerjee\}@rug.nl}}

\begin{document}
\maketitle

\begin{abstract}
The areal footprint of memristors is a key consideration in material-based neuromorophic computing and large-scale architecture integration. Electronic transport in the most widely investigated memristive devices is mediated by filaments, posing a challenge to their scalability in architecture implementation. Here we present a compelling alternative memristive device and demonstrate that areal downscaling leads to enhancement in memristive memory window, while maintaining analogue behavior, contrary to expectations. Our device designs directly integrated on semiconducting Nb-SrTiO$_3$ allows leveraging electric field effects at edges, increasing the dynamic range in smaller devices. Our findings are substantiated by studying the microscopic nature of switching using scanning transmission electron microscopy, in different resistive states, revealing an interfacial layer whose physical extent is influenced by applied electric fields. The ability of Nb-SrTiO$_3$ memristors to satisfy hardware and software requirements with downscaling, while significantly enhancing memristive functionalities, makes them strong contenders for non-von Neumann computing, beyond CMOS.
\end{abstract}


\section{Introduction}
\begin{figure}[b]
    \centering
    \includegraphics[width=\textwidth]{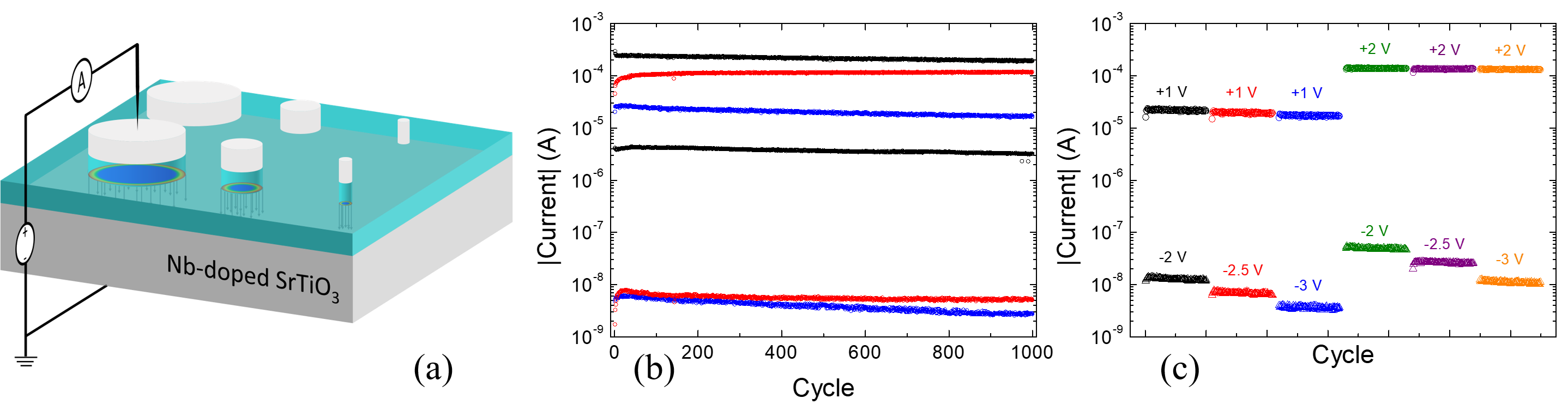}
    \caption{\textbf{State stability and multilevel memristive operation.} \textbf{a} Schematic of the fabricated devices on Nb-doped SrTiO$_3$, electrical connections. Black lines are used to represent the varying overall electric fields acting over each area. The field strengths at the interface are also indicated by a color gradient, showing the fields are weakest in the central area (blue) and strongest around the perimeter (red). \textbf{b} Current read at +0.3 V for device sizes of 100 \textmugreek m (black), 10 \textmugreek m (blue) and 1 \textmugreek m (red). \textbf{c} Current read at 0.3 V after switching between a SET voltage of +1 V (black, red and blue) or +2 V (green, purple and orange) and a RESET voltage of -2 V (black and green), -2.5 V (red and purple) or -3 V (blue and orange). Each combination was repeated over 100 cycles.}
    \label{figure1}
\end{figure}
The growing demand for applications such as artificial intelligence and the \textit{Internet of Things} has given rise to critical challenges in the storage and processing of big data using existing computational architectures \cite{hamdioui2015}. The currently employed von Neumann architecture, using complementary metal-oxide-semiconductor (CMOS) hardware, suffers from limited transmission speed \cite{wilkes1995,mckee2004,horowitz2014} due to a memory throughput bottleneck as well as energy inefficiency and limited scalability \cite{horowitz2014,lahiri2004,hamdioui2013}. Moving away from CMOS technology, towards logic-in-memory chips would alleviate some of the above issues but requires us to massively rethink every aspect of computing \cite{manipatruni2018}. The first step towards this is identifying novel materials and devices with suitable physical properties. Resistive switching devices, or \textit{memristors}, are one such class of devices where the resistance can be switched between several states. Reported in different ionic materials, they are distinguished by the switching mechanism as either occurring through the material bulk between two electrodes or interface-type where switching takes place in a localized region underneath the area of the electrodes \cite{sawa2008}. Their ability to co-locate memory and computation, and exhibit characteristics absent in digital computing makes them important for novel computing approaches. Given the robust way in which the human brain is able to process large amounts of data with remarkably low power, it is unsurprising that it serves as a source of inspiration to the development of computing beyond using CMOS. As the brain utilizes a vast network, downscaling memristive devices is a crucial area of research to develop large scale neuromorphic systems.

For this material-driven research, the areal footprint in unconventional computing architectures that seek to integrate in-memory computing devices such as memristors is a prime consideration. Considerable research has been devoted to this in the realm of non-volatile conventional filamentary devices. The challenges in their implementation in such novel architectures, besides the requirement for unfavourable electroforming processes, lie in their switching endurance \cite{zhang2016}, and their efficacy to exhibit discernible analogue resistance states. Memristive devices that exhibit more than two stable states also greatly enhance integration density because each device can store multiple data bits in an analogue manner.

In valence change memristors, where switching originates from filaments, such behavior is observed in large areal dimensions but is lost when devices are downscaled and conduction is mediated by a single nanoscale filament causing an abrupt transition between the two resistance states \cite{Ang2019}.
Further, the effects of Joule heating on filaments are an important consideration as devices shrink; Joule heating can cause a wide distribution of switching voltages and endurance deterioration. These limitations in device stability, endurance and associated enhanced power of operation are major roadblocks in the successful implementation of filamentary devices in large scale architectures.

Memristive devices have the potential to be integrated in large scale architectures, for which they should exhibit large memory windows, high endurance and low variability \cite{lanza2022}. Herein the areal switching mechanism is a strong contender. A model system in which this mechanism is dominant is Schottky contacts on Nb-doped SrTiO$_3$ (Nb:STO), formed at the interface with a high work function metal. It is widely accepted that in these material systems it is not the bulk of the device, but an area close to the interface that is responsible for the switching, a more detailed discussion on the proposed mechanisms is presented in Supporting Information section S3.
\\
Distinguishing Nb:STO from conventional semiconductors such as Si, widely used in conventional architectures, is its dielectric permittivity which is comparatively large (300) and is strongly dependent on electric field. This property extends the parameter space for designing functionality: electric fields can be used to tune the barrier height and width relevant for memristive device design.
We have previously shown that such Schottky contacts form robust memristors, exhibiting non-linear transport and continuous conductance modulation \cite{goossens2018}, and that their behavior can be described by a power-law which can be successfully implemented as a learning algorithm \cite{tiotto2020}.
However, for the applicability of Nb:STO-based memristors as hardware elements for non-von Neumann computing architecture beyond CMOS, the focus should be on establishing their memristive performance with device miniaturization, which has not been shown on such semiconducting platforms. 
In this work, we demonstrate that memristive devices of Co Schottky contacts on Nb:STO exhibit an increase in the analogue memristive memory window in devices down to 1 \textmugreek m, contrary to expectations.
Ionic defects are at the heart of memristive behavior, hence one of the following two scenarios is expected. For a homogeneous areal mechanism, the current density will scale with device area so that the device resistance in both the high resistance state (HRS) and the low resistance state (LRS) scales with the electrode size, but the ratio between them is area independent. Alternatively, the resistance window can be severely reduced or even vanish with downscaling due to insufficient ionic defects. However, we observe an enhancement in the memory window as the device area is reduced, with minimal device-to-device variation, an unforeseen finding.
\\
To understand the microscopic nature of the switching, we conducted scanning transmission electron microscopy (STEM) on virgin samples and on samples subjected to either a positive (SET) or negative (RESET) voltage . Using integrated differential phase contrast (iDPC) we image oxygen atomic columns next to the heavy metal atomic columns. Virgin samples show the existence of a layer near the interface with neither the perovskite structure of the substrate nor that of the Co electrode. Applying a bias across the interface results in oxygen vacancy movement, which is a key factor controlling the resistance states. 
These new revelations are consolidated with a mathematical model describing the kinetics of trapping and de-trapping in dielectric materials and relates experimental results to the effective trapping density. Surprisingly, this is found to be larger for smaller junctions, suggesting that an increase in the density of traps is responsible for the increased resistance ratio and attributed to inhomogeneous distribution of the electric field due to device edges.

These memristive devices, integrated directly on a semiconducting platform, demonstrate multistate analogue switching with remarkably high memory windows with downscaling, as well as high endurance and low device and cycle variation down to the smallest devices. 
Their ability to meet both hardware and software requirements for unconventional computing, make Nb:STO memristors strong material contenders for physical computing beyond CMOS.

\section{Results}
\subsection{Electrical Characterization}
Figure \ref{figure1}a shows a schematic of the device structure used for the electrical measurements. An array of circular Co electrodes of varying sizes are fabricated on a semiconducting Nb:STO single crystalline substrate. The bottom of the substrate serves as a back contact for the devices. The top electrodes were patterned by a two-step electron lithography process using aluminium oxide as an insulation layer to define the contact areas and to prevent electronic cross talk.

\begin{figure}[t]
    \centering
    \includegraphics[width=\textwidth]{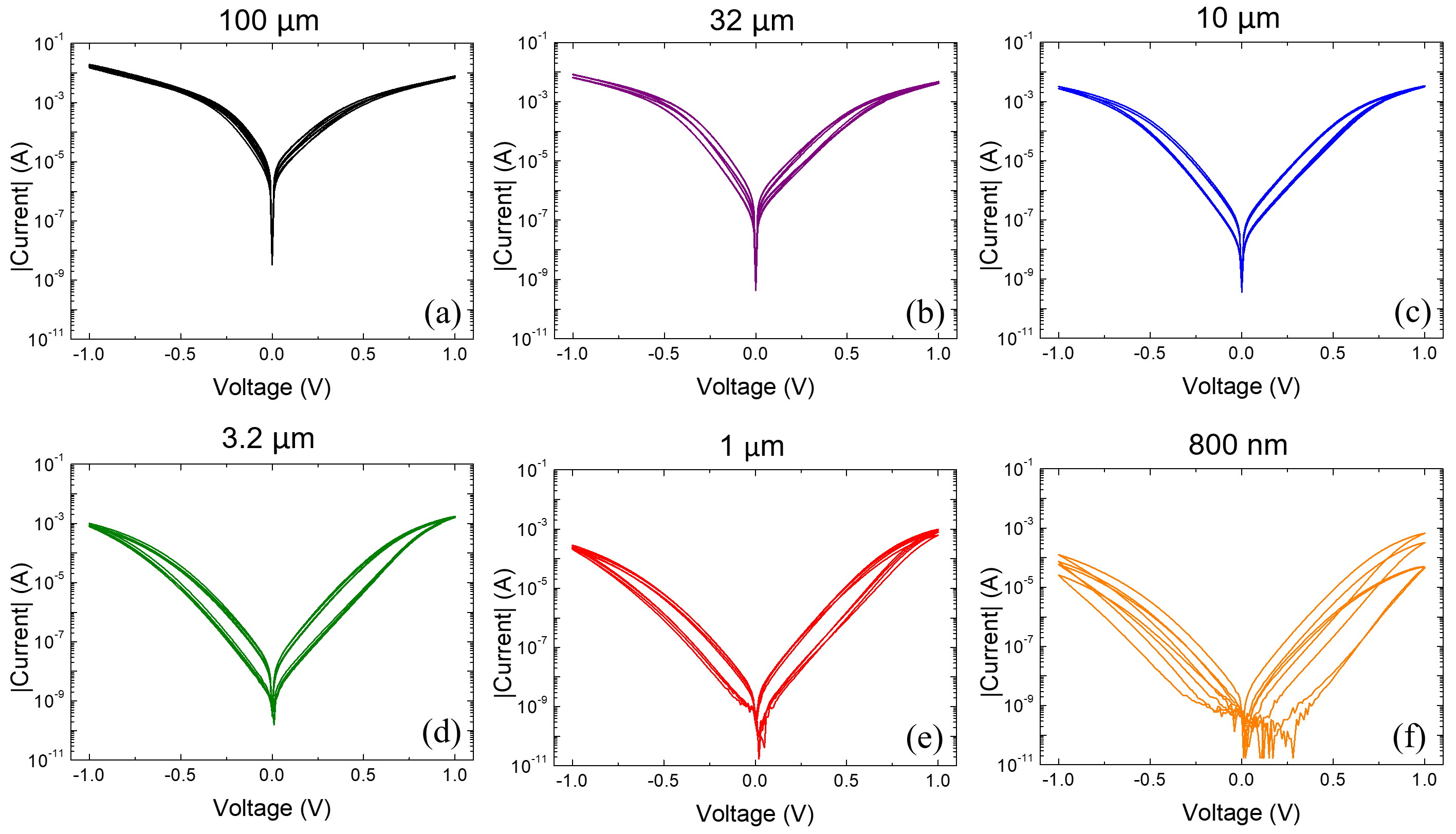}
    \caption{\textbf{Characterization of memristive devices in the virgin state.} Electrical characteristics of virgin devices. The compliance current was fixed at 100 mA for all measurements. Results are shown for four devices of each area in \textbf{a}-\textbf{f}.}
    \label{figure2}
\end{figure}
After fabrication, we performed small range voltage sweeps to characterize the virgin states of each device on a chip. The results for devices with radial dimension from 100 \textmugreek m to 800 nm are shown in Fig. \ref{figure2}, where each sweep followed a voltage sequence from 0 to +1 V to -1 V and back to 0 V. We show four devices of each area, which are plotted in Fig. \ref{figure2}a-f.
\\
The current magnitudes for different devices of the same area show no significant differences down to 1 \textmugreek m, indicating device-to-device variations are minimal. Establishing this is important as this signifies the sole influence of device area in determining the resistance ratio and rules out contributions from device-to-device variation. The 800 nm devices show a greater degree of variation; this is likely due to small differences in their areas and edges arising from the fabrication process and not inherent to the material or due to device fallibility.
No significant differences in the current densities at low bias values are found in the virgin state, confirming that the entire device area contributes to the charge transport (Supporting Information Fig. \ref{fig:Figs1}).
For all the devices, the current gradually increases and exhibits a small hysteretic effect from the virgin state, indicating that no forming step is required. 
\begin{figure}[t]
    \centering
    \includegraphics[width=\textwidth]{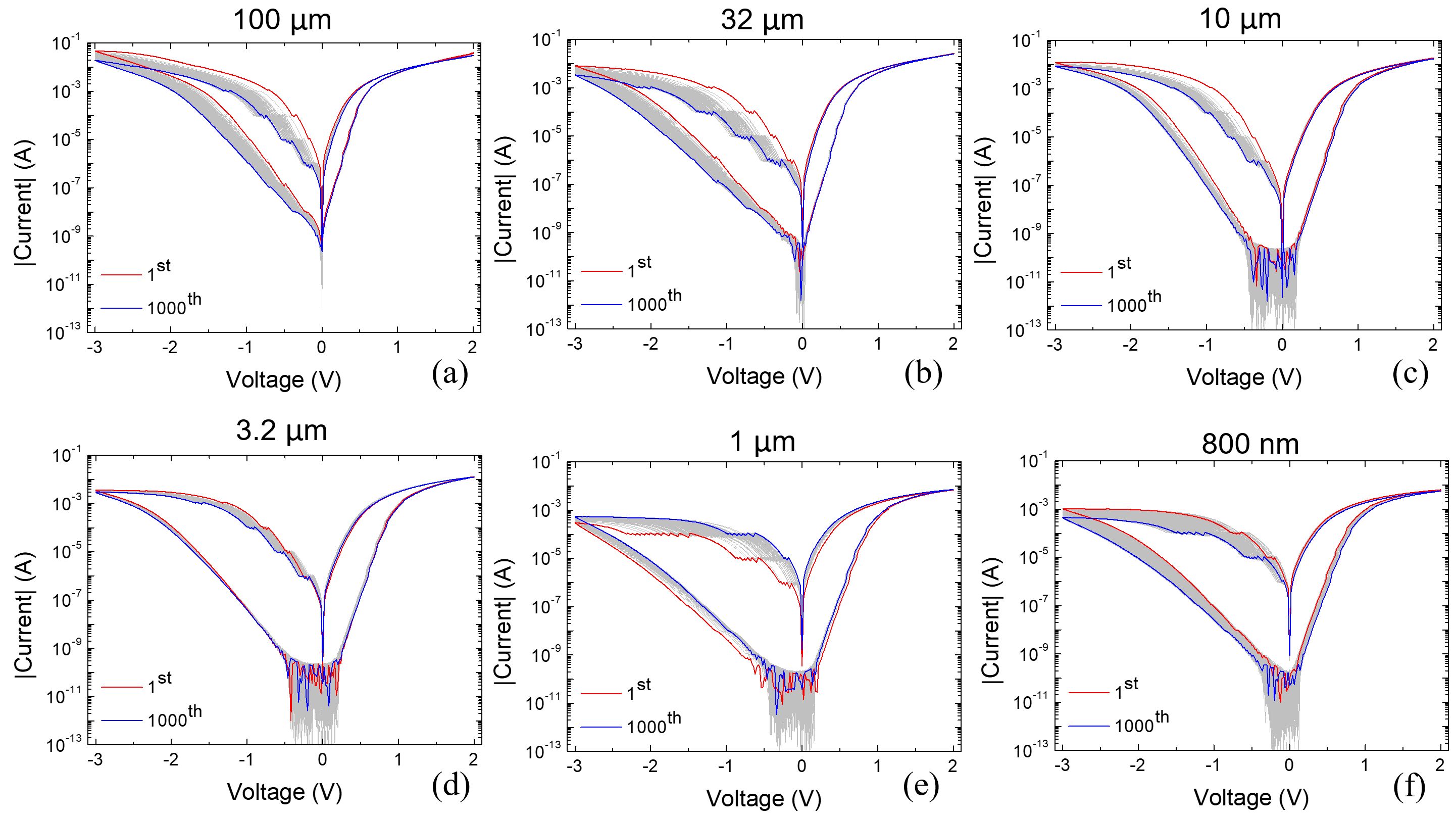}
    \caption{\textbf{Resistance ratio, cycling endurance and state stability. }\textbf{a}-\textbf{f} show 1000 consecutive current-voltage sweeps from +2 V to -3 V to +2 V at a rate of 1.52 Vs$^{-1}$ for devices of 100 \textmugreek m down to 800 nm. Starting from a SET voltage of +2 V, each device is in an LRS, represented by the upper branch reaching the RESET voltage of -3 V and sweeping back, the devices are switched to an HRS (represented by the lower branch.}
    \label{figure3}
\end{figure}

Figure \ref{figure3}a-f shows 1000 consecutive current-voltage (I-V) sweeps of these devices. Starting from a SET voltage of +2 V, each device is in an LRS, represented by the upper branches. After reaching the RESET voltage of -3 V and sweeping back, the devices are switched to an HRS (represented by the lower branches). In all device areas both the SET and RESET operation remain continuous, indicating the resistive switching retains its analog nature when downscaling. The cycling endurance was measured for over 10$^5$ switching cycles without device failure, illustrating an endurance of $>$10$^5$. The current in the HRSs scales approximately with area at low bias values, while the low resistance current, is less closely correlated to the area. As a result, the resistance window increases with decreasing device area in both forward and reverse bias. Figure \ref{figure1}b and Supporting Information Fig. \ref{fig:Figs2} show the current and current density at a low read voltage of 0.3 V, respectively. Minimal cycle-to-cycle variations at low reading voltages are found with reproducible switching between clearly distinguishable states without degradation in device performance. This also establishes the low power operation of these devices after downscaling, which is important for memristor operation. As shown in Supporting Information Fig. \ref{fig:Figs3}, the device-to-device variation remains low down to 1 \textmugreek m. The variation in the resistance ratio in the 800 nm devices is larger (Fig. \ref{fig:Figs4}), and will be discussed later.

The SET and RESET transitions are gradual and highly tunable. To demonstrate this, a 1 \textmugreek m device was subjected to voltage sweeps varying between different positive (SET) and negative (RESET) voltages. Figure \ref{figure1}c shows that a wide range of stable states is available at a low read voltage of +0.3 V. The wide dynamic range combined with the large number of distinct addressable states ensures device reliability and increased memory storage capabilities. Each state maintains a narrow distribution of current values over the 100 cycles shown, reiterating the stability of the switching process.

\begin{figure}[t]
    \centering
    \includegraphics[width=\textwidth]{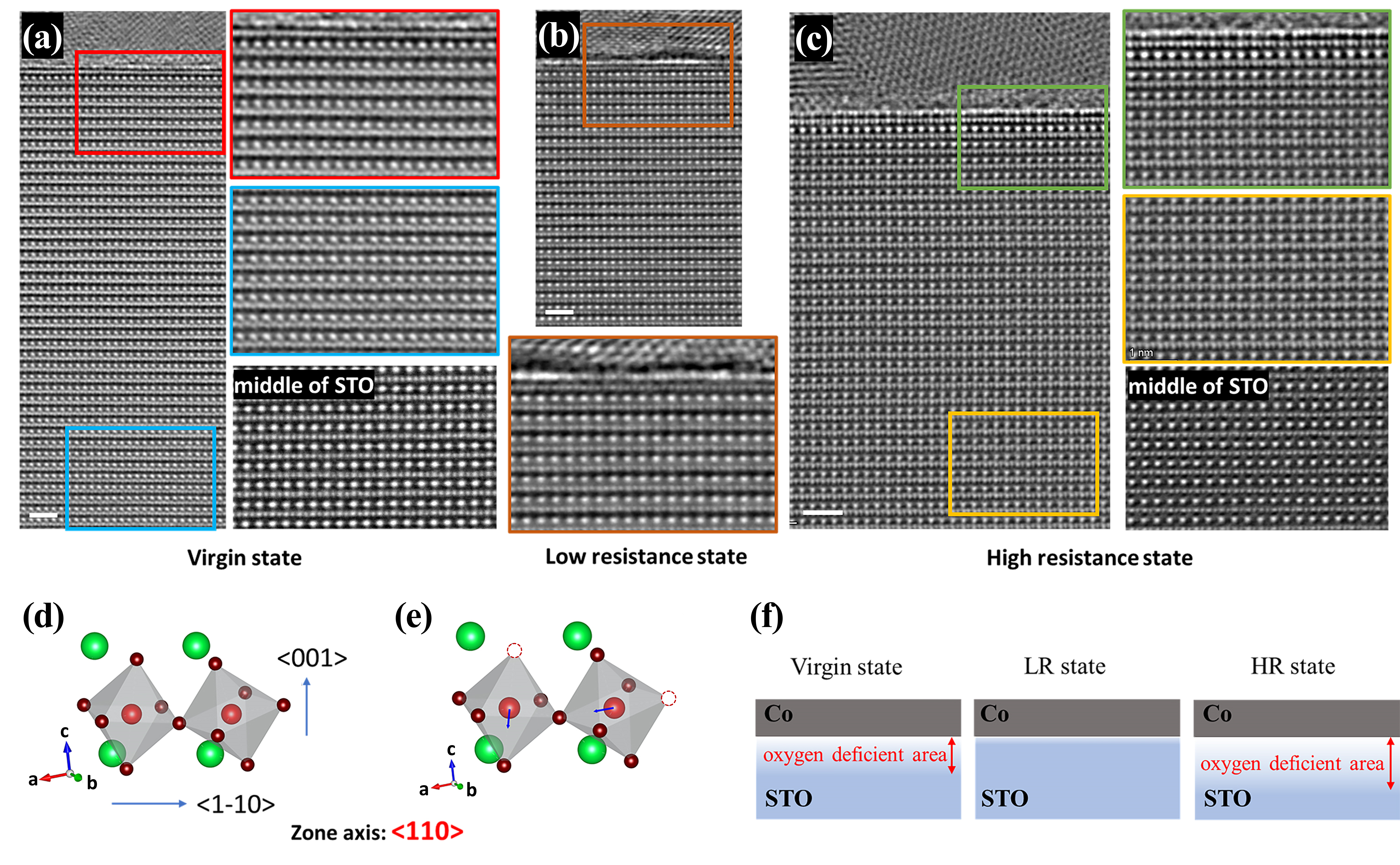}
    \caption{\textbf{Visualization of oxygen vacancy migration using STEM.} iDPC-STEM images of Co/Nb:STO samples in \textbf{a} the virgin (unbiased) state, \textbf{b} the LRS and (\textbf{c} the HRS, highlighting the structure close and far from the interface. The perovskite unit cell of STO, showing Sr in green, O in dark red and Ti in light red, viewed along the $<$110$>$ in \textbf{d} the pristine state and (\textbf{e} with oxygen vacancies. The deficiency of O causes Ti atoms to move away from the vacancies as shown by the arrows. \textbf{f} shows a schematic representation of how the interfacial layer is affected by biasing.}
    \label{figure4}
\end{figure}
\subsection{Scanning Transmission Electron Microscopy}
A microscopy study of the Schottky interface was carried out using STEM. Figure \ref{figure4} shows atomic resolution cross-section STEM-integrated Differential Phase Contrast (iDPC) images of the Co/Nb:STO interface for samples in the unbiased virgin condition (Fig. \ref{figure4}a), the LRS state (Fig. \ref{figure4}b) and the HRS state (Fig. \ref{figure4}c). To image lighter oxygen atoms, integrated into a matrix with heavier Sr and Ti atoms, we utilized STEM-integrated Differential Phase Contrast (iDPC) instead of the more commonly employed STEM-High-angle annular dark-field (HAADF) imaging technique.\cite{lazic2016,degraaf2020}.
The STEM images in Fig. \ref{figure4}a show that, apart from a thin interfacial region, the bulk STO consists of a cubic perovskite lattice and no defects are observable. All images taken within the bulk did not show any dislocation and possessed the expected perovskite structure as shown in Fig. \ref{figure4}d.
However, the structure close to the interface deviates from this perovskite structure and is deficient in oxygen. The migration of oxygen ions near the interface towards Co causes positively charged Ti ions to be displaced so that they no longer sit equidistantly from the Sr ions along $<$001$>$. Figure \ref{figure4}e illustrates how the loss of O ions gives rise to Ti displacements along the $<$001$>$ direction away from the interface as well as along $<$1-10$>$ (see Supporting Information Fig. \ref{fig:Figs7}) and is similar to what was reported in ref. \cite{nukala2021} in La$_{0.67}$Sr$_{0.33}$MnO$_3$/Hf$_{0.5}$Zr$_{0.5}$O$_2$. We believe the creation of this thin layer to be related to the formation of a Schottky barrier. The analysis for a non-memristive interface with Ti contacts can be found in Supporting Information Fig. \ref{fig:Figs6}.

Figure \ref{figure4}b shows analogous results to Fig. \ref{figure4}a, but now for the sample switched to the LRS, representing the upper branch in Fig. \ref{figure3}, after the application of a positive bias voltage of 2 V. Comparing the two figures shows that in the LRS state the extent of the interfacial layer has decreased. This suggests that under the influence of a positive voltage, the labile bonds between O and interfacial Co atoms are broken and oxygen moves back into the STO substrate.
A negative bias voltage of -3 V (corresponding to the lower branch in Fig. \ref{figure3}), on the other hand, causes oxygen to move from STO to cobalt causing the formation of CoO and more oxygen vacancies in the STO, highlighted by a larger region over which Ti ions are displaced (see Fig. \ref{figure4}c). This indicates that the formation of the CoO switches the sample to the HRS state. 
It has been shown \cite{sung2013two,lee2014anomalous} that the oxygen vacancy distribution inside the system will determine how the oxygen vacancies are affected by the applied voltage. The formation of an oxygen deficient interfacial layer confirms that in these samples the oxygen vacancies are concentrated near the interface. In this case, it is expected that the application of a positive voltage will cause oxygen vacancies to be repelled from the interface while a negative voltage will cause oxygen vacancies to be attracted to the interface, consistent with our findings.
After removing the voltage, the interfacial layer did not reform over time, suggesting the presence of an oxygen-migration blocking layer. 
These results are summarized in Fig. \ref{figure4}f.

Our results directly confirm the existence of a homogeneous oxygen deficient layer at the interface. The homogeneous nature of the defect state layer ensures ionic defects are retained with downscaling.
We furthermore show that the physical extent of the layer is reduced or extended when a positive or negative voltage is applied respectively.
Although the uniform nature of the ionic contribution to switching is now verified, this does not explain the origin of the unexpected enhancement of the resistance window with downscaling. This we discuss next by considering the trapping of electronic charges at oxygen vacancy sites.

\subsection{Model}
\begin{figure}[t]
    \centering
    \includegraphics[width=\textwidth]{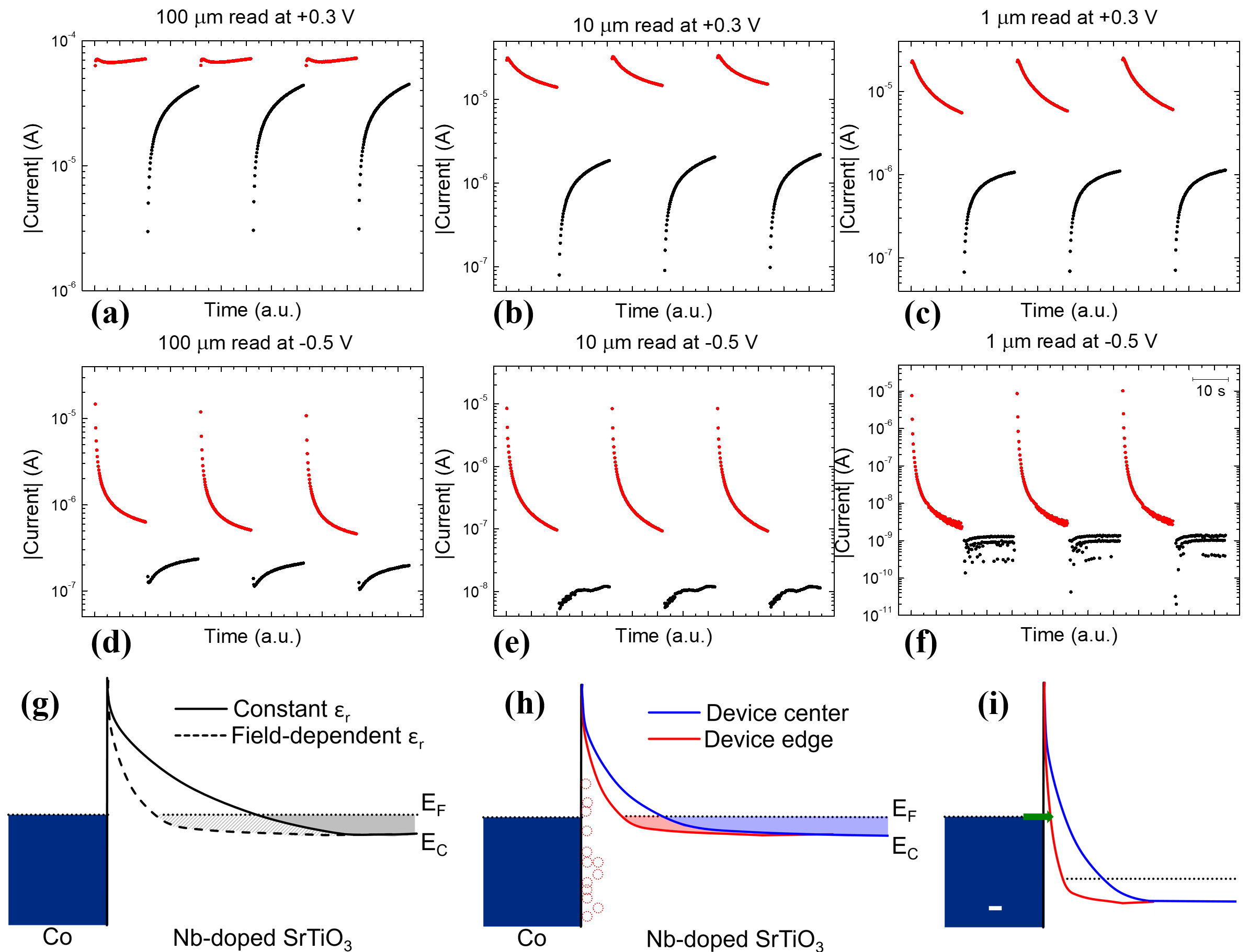}
    \caption{\textbf{Trapping dynamics and Schottky interface energy landscapes. }Retention characteristics of differently sized devices read at +0.3 V \textbf{a}-\textbf{c}) and -0.5 V (\textbf{d}-\textbf{f}) after a SET voltage of +2 V (red) or -3 V (black). \textbf{g} shows the energy landscape of a Schottky interface in equilibrium when the dielectric constant does not depend on electric field (solid line) and when the dielectric constant is field-dependent (dashed line). $E_F$ and $E_C$ are the Fermi level and conduction band respectively. The energy landscapes at the center and edge of a device are compared in \textbf{h} in equilibrium and \textbf{i} in reverse bias. Red circles represent oxygen vacancy states and the green arrow indicates electron tunneling.}
    \label{figure5}
\end{figure}

In order to understand how the electrical properties of the devices are influenced by these oxygen vacancies, we consider the interaction between electrons and defect states. This interaction is most strongly evidenced by the retention characteristics, which have a slow decaying component. This behavior is caused by the detrapping of charges. It has been shown that this occurs over long timescales and the different states will remain clearly distinguishable for long time periods of hours and that the retention time is tunable by the applied stimuli \cite{goossens2018}. We utilized short voltage pulses to measure the retention characteristics of each device in both an HRS and LRS. This was done by applying alternating SET and RESET pulses of +2 V and -3 V respectively, and reading the small-signal current at either +0.3 V or -0.5 V after each writing event. The state retention characteristics of the different devices are shown in Fig. \ref{figure5} for the LRS (red) and HRS (black). Over time, the current in both states tends to an intermediate value. For the LRS, the rate of change follows a power law that is commonly observed for charge trapping under bias in high-$\kappa$ dielectrics, referred to as the Curie-von Schweidler law. 

This law describes a non-Debye type relaxation in dielectrics. Empirical evidence of this behavior is seen in a wide variety of materials, but the precise physical origin remains unclear. 
Here we consider the effect of injected electrons becoming trapped in defects states within the dielectric. The space charge generated by these trapped electrons lowers the electric field, in turn reducing the flow of current through the dielectric. 
In this case the trapping rate can be expressed as:
\begin{equation}
    \frac{dn}{dt}=n_0\sigma\frac{Jv_{th}}{qv_d}e^{-\frac{nh}{V}}
\end{equation}
where $n_0$ is the maximum number of traps available, $J/q$ is the net flux density, $v_{th}$ and $v_d$ are the thermal and drift velocities respectively, and $\sigma$ is capture cross-section. 
Solving this equation yields the following expression for $n$:
\begin{equation}
\label{n}
    n=\frac{V}{h}\ln\Big(\frac{Q}{Q^*}+1\Big)
\end{equation}
where $Q=\int Jdt$ is the total injected charge and
\begin{equation}
    Q^*=\frac{Vv_dq}{n_0hv_{th}\sigma}
\end{equation}
Expressing the current as $J=J_st^{-\alpha}$ and extending this analysis results in:
\begin{equation}
\label{proofmain}
    \frac{\alpha}{1-\alpha}\ln(J_s)\approx mE_{ap}+n
\end{equation}
where $m$ is a constant.
 
We can also directly relate the trapping rate to the current.
$Q_T$ represents the charge that is trapped when charge $Q$ is injected into the dielectric. The ratio $\frac{dQ_T}{dQ}$ is a function of current.
The current can be written as:
\begin{equation}
    J=J_s\frac{t}{t_0}^{-1/(\alpha+1)}
\end{equation}
where $\alpha\geq$0 and $J_s$ depends on the transport mechanism. For conduction following an exponential relation:
\begin{equation}
\label{exp}
    J_s\propto e^{\frac{(1-1/(\alpha+1))V}{V_0}}
\end{equation}
Here, $V_0$ is a constant. The full derivation is shown in Supporting Information Section S1 and Fig. \ref{fig:Figs8}, and is also extended to show that it holds for other transport mechanisms.

Equations \ref{proofmain} and \ref{exp} serve as a direct mathematical proof that the exponent $\alpha$ in the power law is related to the effective trap density or capacity of the dielectric to trap electrons. This derivation is applicable to a wider range of systems, irrespective of the choice of dielectric material.
\begin{table}[t]
\centering
$\rotatebox[origin=c]{90}{area}%
\left\downarrow
\begin{tabular}{c|cc}
\hline
\multicolumn{1}{l|}{}                        & \multicolumn{2}{c}{\textbf{$\lvert$Exponent$\rvert$}}                                \\
\multicolumn{1}{l|}{\textbf{Radius (\textmugreek m)}} & \multicolumn{1}{c|}{\textbf{Read at +0.3 V}} & \textbf{Read at -0.5 V} \\ \hline
1                                           & \multicolumn{1}{c|}{0.85$\pm$0.03}         & 2.17$\pm$0.02         \\
10                                           & \multicolumn{1}{c|}{0.47$\pm$0.02}         & 0.987$\pm$0.002         \\
100                                          & \multicolumn{1}{c|}{0.041$\pm$0.004}         & 0.626$\pm$0.007         \\ \hline
\end{tabular}
\right\uparrow
\rotatebox[origin=c]{90}{trapping density}$
\caption{Magnitude of exponents, $\alpha$, extracted by fitting a power-law to the low resistance states in the graphs in Fig. \ref{figure5}.}
\label{table}
\end{table}
In Table \ref{table} we show the LRS exponents, $\alpha$ for each device. Larger values are observed for smaller devices indicating that the trap density is higher in the smallest device compared to the larger device.
\section{Discussion}
While this model provides a clear correlation between trapping density and device area, it does not give information about the traps; we implicitly take all traps to be of the same kind, while in reality, the nature of traps can vary greatly. 
The trapping rate can depend on the spatial location of the traps and new traps can be generated via defect migration. For a more precise picture of the mechanism, we need to consider a distribution of traps with respect to their location within the dielectric. 
Evidenced by the STEM study, oxygen vacancies are the most important class of trapping defects to consider. They are abundantly present in SrTiO$_3$ due to their low formation (0.51 eV\cite{desouza2012}) and migration (0.62 eV\cite{iglesias2017}) enthalpies and their locations within the energy landscape are well documented\cite{freysoldt2014}. 

From the discussion above, it is clear that the energy landscape of these Schottky junctions is far more complex than is captured by the most commonly used models that are based solely on parameters of the individual materials forming the contact \cite{schottky1939,mott1939}. Transport through these junctions is usually described by the thermionic emission equation, which includes an ideality factor accounting for the deviating transport from this ideal diode equation. This model furthermore does not consider that the interfacial area is not spatially homogeneous and that in devices of finite areas the boundary of the device will be relevant. In particular, it is known that near the edges crowding of the field lines leads to an enhancement in the field strength which can decrease the barrier width \cite{rhoderick1982,willis1990}. 
This is supported by the results of the finite element simulations in Supporting Information Fig. \ref{fig:Figs11} and \ref{fig:Figs12}, showing a significant enhancement in the electric field around the edge and when downscaling. 
From the simulations it is evident that there is still a clear field gradient in the 1 \textmugreek m devices, indicating that a further increase in ratio with downscaling can be expected, and the areal field shows no apparent saturation till around 10 nm (Fig. \ref{fig:Figs13}).

The observed enhancement is especially important in Nb:STO-based memristive devices as the dielectric constant of the substrate strongly depends on electric fields \cite{barrett1952dielectric,van1995field}. This will further alter the potential landscape of the Schottky interface in such memristive devices. In particular, the dielectric permittivity of Nb:STO rapidly decreases in the presence of large electric fields which results in a decrease in the effective Schottky barrier width as illustrated in Fig. \ref{figure5}g. Consequently, a large reduction in the barrier width is expected to occur near the device edges (Fig. \ref{figure5}h). It has also been shown that an electric field can modify the defect states and significantly affect trapping parameters\cite{dussel1966electric}.

Given that the charge transport is governed by the potential landscape, this will hugely impact the measured current, pictured in Fig. \ref{figure5}i. Tunneling through the barrier will be enhanced near the device edges leading to a larger current near the device perimeter. This will be especially important in the LRS where the interface is depleted of trapped charges and the Schottky barrier is narrower, leading to more tunneling \cite{mikheev2014,goossens2018}. 

Transport across the interface is comprised of thermionic emission and tunneling. The thermionic current density is expected to be independent of area and is the dominant mechanism in the HRS at low bias voltages, giving rise to the decreasing current in the HRS around zero with downscaling observed in Fig. \ref{figure3}.
At higher voltage values, however, tunneling will also contribute to the current; the tunneling current density will increase with decreasing area. In Fig. \ref{figure1}b, the current is read at +0.3 V where we expect both thermionic emission and tunneling to contribute to transport, giving rise to similar currents measured for the 10 and 1 \textmugreek m devices in the HRS. The tunneling contribution increases in the LRS, especially in smaller devices due to the larger electric fields, resulting in the observed increase in current density with reducing area.

By applying a potential over the Schottky barrier, the Fermi level is shifted such that tunneling electrons sample different oxygen vacancy energy levels. As the reverse bias voltage is increased, electrons are gradually exposed to larger ranges of states in which they can become trapped. In addition, in reverse bias, the electric field at the interface becomes larger leading to a reduction in the dielectric constant and a corresponding decrease of the Schottky barrier widths. This decrease in width will be more pronounced in regions closer to the edge due to the local field enhancement. As a result of the narrower barrier, electron-electron scattering will be reduced and the trap states will act as the main barrier for transport. The stronger edge field may additionally facilitate the migration of oxygen vacancies resulting in a higher number of vacancies accumulating around the perimeter. Consequently, the trapping efficiency will be greater near the edge than in the center. This is a unique effect enabled by the electric field control of the dielectric permittivity, does not occur in conventional semiconductors and is relevant for Nb:STO memristive device design.

We can express the area and perimeter of a device with radius $r$ as $A=\pi r^2$ and $p=2\pi r$ respectively. The ratio of the perimeter to area:
\begin{equation}
\label{ratio}
    \frac{p}{A}=\frac{2\pi r}{\pi r^2}=\frac{2}{r}
\end{equation}
indicates that the edge effects become more dominant as the device area is reduced. As a result, current flow at the perimeter will constitute a larger percentage to the overall transport behavior in smaller devices. This explains the enhanced current densities observed when downscaling after applying large bias voltages as well as the larger effective trapping densities for smaller devices. 
Specifically, this field enhancement around the device edges gives rise to an increase in the dynamic range in smaller devices, and explains the unexpected resistance window scaling.
\section{Conclusions}
As a first demonstration of exploiting edge effect related additional electric fields, our work successfully demonstrates the ability to increase the resistance window by device miniaturization of interface memristors from 100 \textmugreek m down to 1 \textmugreek m, contrary to expectations, with exceptional robustness to device-to-device and cycle variability. Scanning transmission electron microscopy images taken in the virgin, high and low resistance states prove the existence of a homogeneous interfacial layer, deficient in oxygen, whose physical extent is influenced by applying an electric field. This, however, does not explain the enhancement in the resistance window with device downscaling. A model describing the interaction of electrons with oxygen vacancy trap states shows an increase in the effective trapping density with downscaling. The advantage of direct integration of devices on a semiconducting platform of Nb-doped SrTiO$_3$ allows for the locally enhanced fields to controllably tune the interfacial energy landscape at the interface, leading to a greater contribution of edge effects in smaller devices as confirmed by finite element simulations. With rapid advances made in the palette of materials and devices available for neuromorphic hardware, the thrust now should be in their efficient integration on semiconducting platforms for on-chip applications with substantial reduction in areal footprint. In this, our work provides an encouraging direction. 

\section{Experimental Section}
\subsection{Electrical Device Fabrication} 
We investigated a series of Co/Nb-doped SrTiO$_3$ devices, where the device area was varied across the series over a range spanning five orders of magnitude ranging from 10$^{-12}$ to 10$^{-8}$ m$^2$, with radii between 800 nm and 100 \textmugreek m. The devices were fabricated using Nb-doped SrTiO$_3$ (001) substrates with a doping concentration of 0.1 wt$\%$ from Crystec. SrTiO$_3$ consists of alternating SrO and TiO$_2$ planes along the [001] direction. The as-received substrates have a slight miscut from the exact crystallographic direction and as a result, a mixture of both terminations exists at the surface. It has been shown that the local properties of Schottky barriers grown on the different terminations may differ, hence to minimize the variation of different areas on the substrate a single termination is desired. To ensure that the terminating layer is TiO$_2$, a chemical treatment was carried out with buffered hydrofluoric acid (BHF). A further annealing treatment at 960 $^{\circ}$C in an O$_2$ flow of 300 ccmin$^{-1}$ to facilitate the reorientation of surface atoms to form an atomically flat and straight terraced surface. Atomic force microscopy images were taken at different parts of the substrate and confirmed the existence of uniform terraces. The substrate was then coated with a negative resist (AZ nLOF 2020) and using electron beam lithography circles of different areas were patterned. A thick insulation layer of AlO$_x$ was deposited using electron beam evaporation and lift-off was carried out to define a set of direct contacts to the substrate. By means of a second lithography step with a positive resist (950 K PMMA), square contact pads were defined, each covering a hole and part of the surrounding AlO$_x$: the dimensions of these pads were identical for each device to minimize spurious effects arising from significantly different contact resistances. Co (20 nm) and a capping layer of Au (100 nm) were then deposited using electron beam evaporation in high vacuum ($\sim$10$^{-6}$ Torr).

\subsection{Electrical Characterization} 
Electrical measurements were conducted using probes connected to two remote-sense and switch units (RSU) of a Keysight B1500A Semiconductor Device Parameter Analyzer. During the voltage sweeping measurements, conducted using a sweeping measurement unit (SMU), the bottom of the substrate is held at 0 V while a voltage is applied to the top electrode. Due to the diodic nature of the devices in conjunction with large degrees of resistive switching, the measured currents during a single sweeping measurement span up to 9 orders of magnitude. For this reason, the measurements were performed using auto range for the measured current. The effects of this can be observed in the endurance cycling measurements which were performed at high sweeping rates in the form of plateaus in the current whenever a limit of the SMU range is reached. 

\subsection{Scanning Transmission Electron Microscopy} 
The samples discussed in this work use SrTiO$_3$ (001) substrates with an Nb-doping in place of Ti of 0.1 wt$\%$ from Crystec. The surface was prepared using a chemical treatment with buffered hydrofluoric acid (BHF). Next, the substrates were annealed at 960$^{\circ}$C in an O$_2$ flow of 300 ccmin$^{-1}$.
For STEM samples films were deposited by electron beam evaporation of 20 nm of Co capped with 20 nm of Au and 20 nm of Pt. From this, three types of STEM lamellae were prepared: virgin (unbiased) samples, low resistance state (LRS) samples and high resistance state (HRS) samples. Using a probe station, samples are subjected to bias values of +2 V and -3 V to prepare samples in the LRS and HRS respectively. 
STEM lamellae were extracted from samples along the $<$110$>$ direction using a Helios G4 CX dual beam system with a Ga focused ion beam. The lamellae were thinned to make them transparent to electrons using the focused ion beam. Imaging was carried out using a Thermo Fisher Scientific Themis Z S/TEM system operating at 300 kV.

STEM-High-angle annular dark-field (HAADF) images are most widely used, because they are readily interpretable with atomic columns being bright spots in a dark surrounding, where the brightness of the spots scale with the average atomic number $Z$ ($\sim$Z$^{1.7}$). This technique is well suited to image heavy elements, but lighter elements, such as oxygen, are harder to detect, and cannot be detected properly when integrated into a matrix with much heavier elements (like Sr). Therefore, to gain more insight into the important role played here by the oxygen ions, we utilized here STEM-iDPC instead of STEM-HAADF imaging. This technique uses a four-quadrant annular bright field detector and can be used to acquire the projected local electrostatic potential of the sample (when thin) and has clear advantages over traditional annular bright field (ABF) imaging \cite{lazic2016,degraaf2020}.
\subsection{Simulations}
Finite element modeling of the electric field profile at the interface was carried out using COMSOL Multiphysics.
\subsection{Statistical Analysis}
For the $\lvert$current$\rvert$-voltage graphs, the absolute value of the measured current is taken; to determine the current density, the measured current was divided by the area of the Co contact. The values in Table \ref{table} were derived by iteratively fitting the data in Fig. \ref{figure5}a-f using a power-law equation of the form $I=I_0(t-t_0)^{-\alpha}$ by means of the Levenberg-Marquardt algorithm; the reported errors are the standard errors calculated by this method. The fits are shown in supporting Fig. \ref{fig:Figs9}. The inverse scaling of the exponent and device area was verified for different devices and different reading and SET voltages. Plotting and analysis of electrical measurements was done using OriginPro 8.5. Measurements were repeated on four devices of each area to check reproducibility and validity of results.\\
For STEM images, multiple regions for each one of the three bias conditions were taken to verify the results. The idpc images were filtered by applying a high-pass Gaussian filter using Velox.

\medskip
\textbf{Acknowledgements} \par 
A.G. is supported by the CogniGron Center, University of Groningen. Device fabrication was realized using NanoLab NL facilities. We acknowledge technical support from J. G. Holstein, H. H. de Vries, A. Joshua, T. Schouten, and H. Adema. We thank R. J. E Hueting, P. Nukala, S. de Graaf and T. Kenyon for useful discussions. A.G., D.G, I.B, and T.B. benefited from helpful discussions with the members of the Spintronics of Functional Materials group.

\medskip
\textbf{Conﬂict of Interest} \par The authors declare no conﬂict of interest

\medskip
\textbf{Author Contributions} \par A.G. and T.B. conceived the idea and designed the devices. A.G. and D.G. fabricated devices for electrical measurements and performed electrical measurements, along with I.B.. 
D.G. derived the mathematical model discussed in the manuscript. 
M.A. and A.G. fabricated lamalae for STEM and M.A. took STEM images.
Finite element simulations were done by A.G..
All authors analyzed the data, discussed the results and agreed on their implications. All authors contributed to the preparation of the manuscript. 


\bibliographystyle{unsrt}
\bibliography{arxiv}

\clearpage

\renewcommand{\thepage}{S\arabic{page}} 
\renewcommand{\thesection}{S\arabic{section}}  
\renewcommand{\thetable}{S\arabic{table}}  
\renewcommand{\thefigure}{S\arabic{figure}}
\renewcommand{\theequation}{S\arabic{equation}}
\setcounter{figure}{0} 
\setcounter{table}{0} 
\setcounter{equation}{0} 
\setcounter{page}{1}

\section*{Supplementary Data}

\begin{figure}[H]
	\centering
	\includegraphics[width=0.8\textwidth]{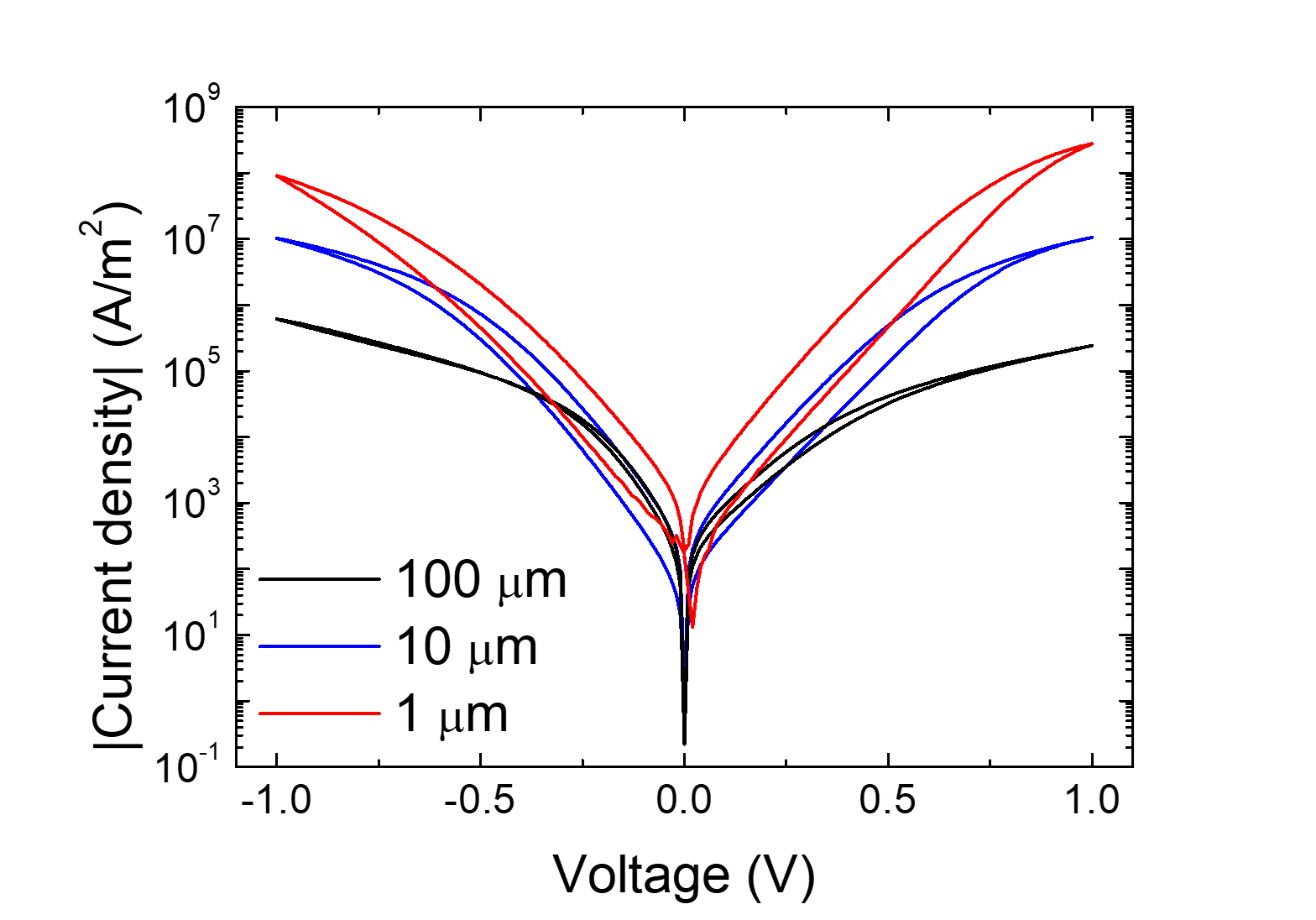}
	\caption{
\textbf{Current densities of virgin devices.} Results are shown for devices of radial dimensions of 100 \textmugreek m (black), 10 \textmugreek m (blue) and 1 \textmugreek m (red).
}
\label{fig:Figs1}
\end{figure}
\begin{figure}[H]
	\centering
	\includegraphics[width=0.7\textwidth]{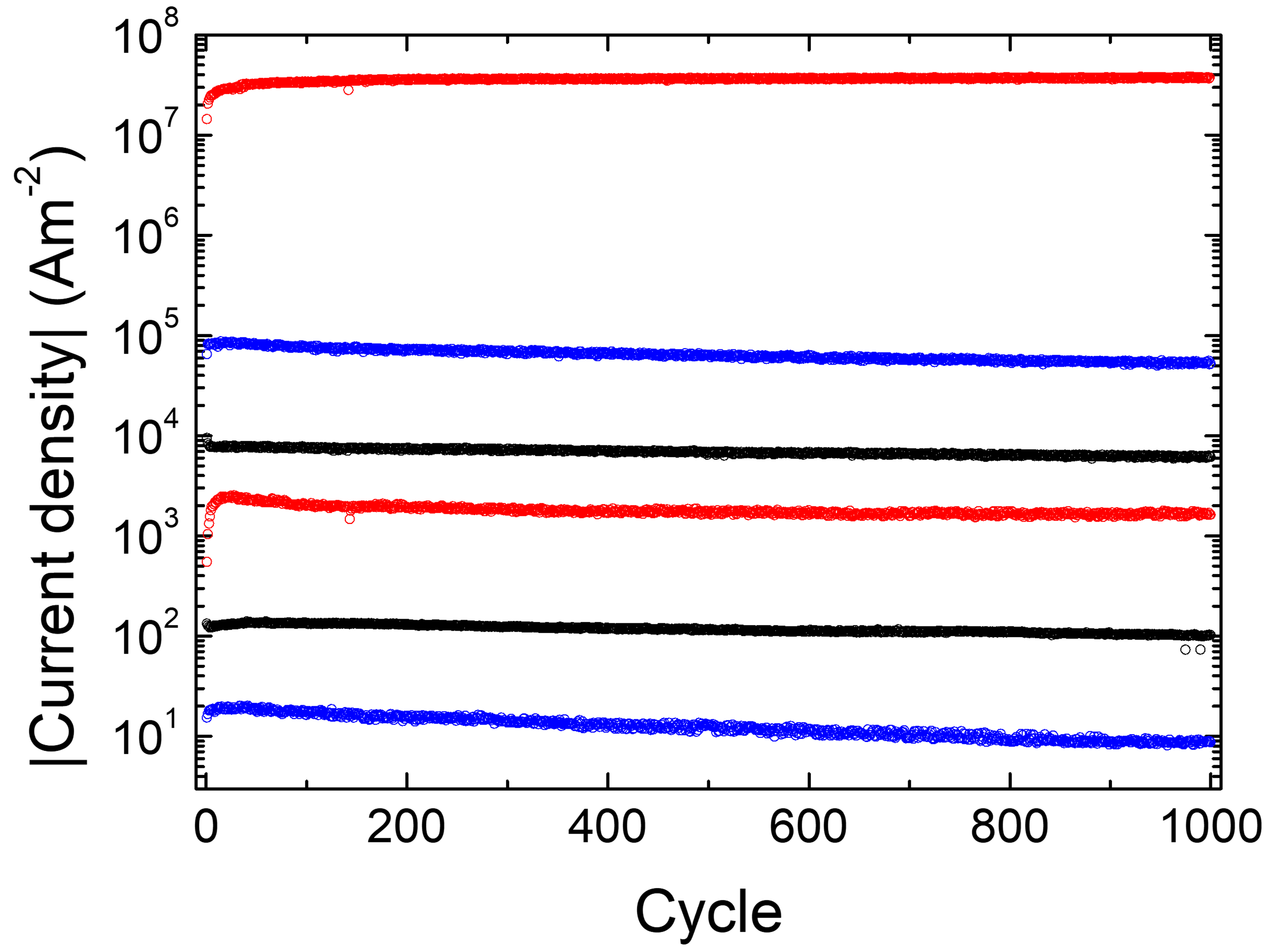}
	\caption{
\textbf{Cycle-to-cycle variation.} The current density read at +0.3 V for device sizes of 100 \textmugreek m (black), 10 \textmugreek m (blue) and 1 \textmugreek m (red).
}
\label{fig:Figs2}
\end{figure}

\begin{figure}[H]
	\centering
	\includegraphics[width=\textwidth]{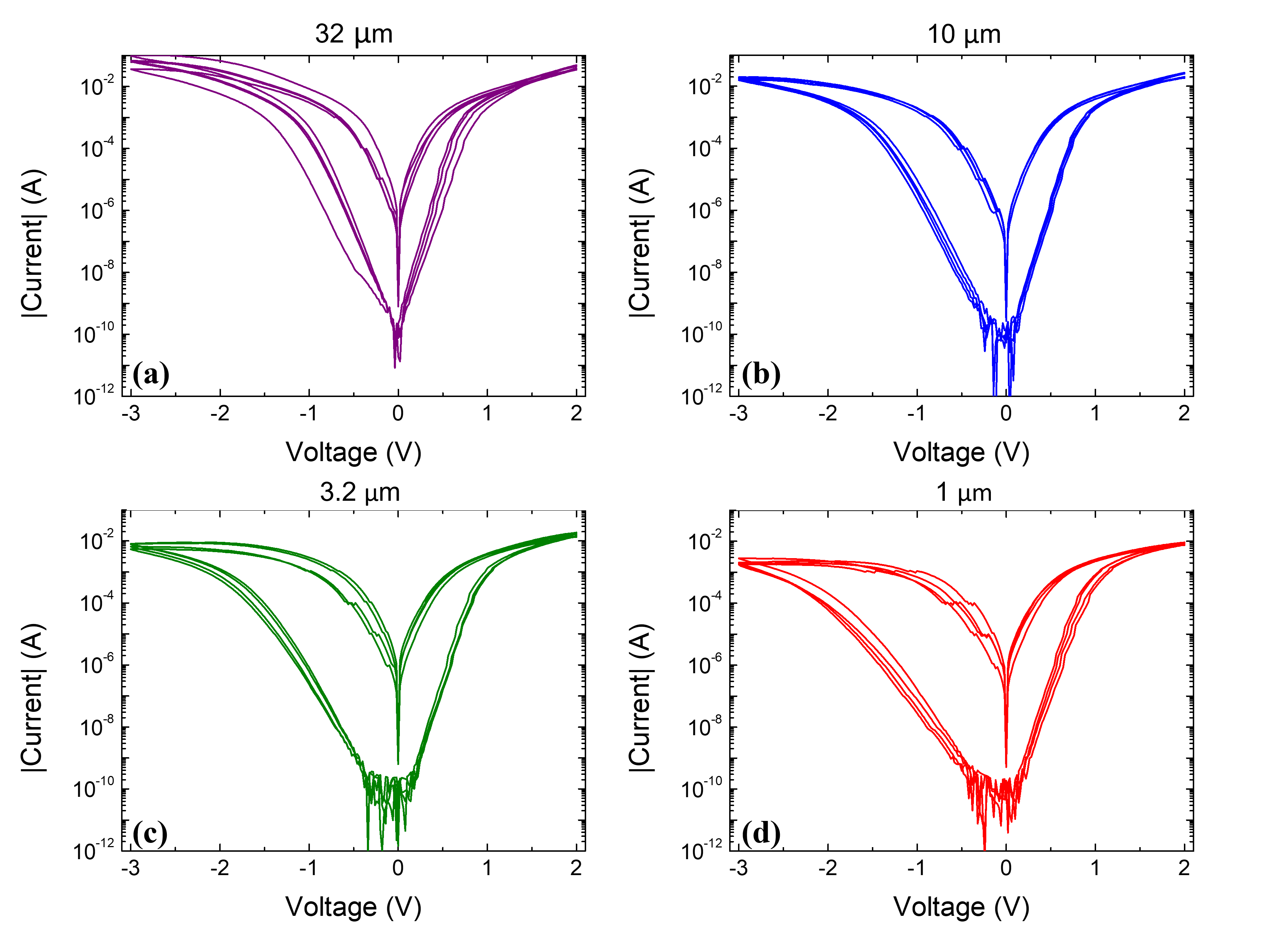}
	\caption{
\textbf{Device-to-device variation.} Current-voltage sweeps from +2 V to -3 V to +2 V at a rate of 1.52 Vs$^{-1}$ measured between a SET voltage of +2 V and a -3 V RESET voltage. Measurements are shown for different devices to demonstrate the low device-to-device variability.
}
\label{fig:Figs3}
\end{figure}
\begin{figure}[H]
	\centering
	\includegraphics[width=0.6\textwidth]{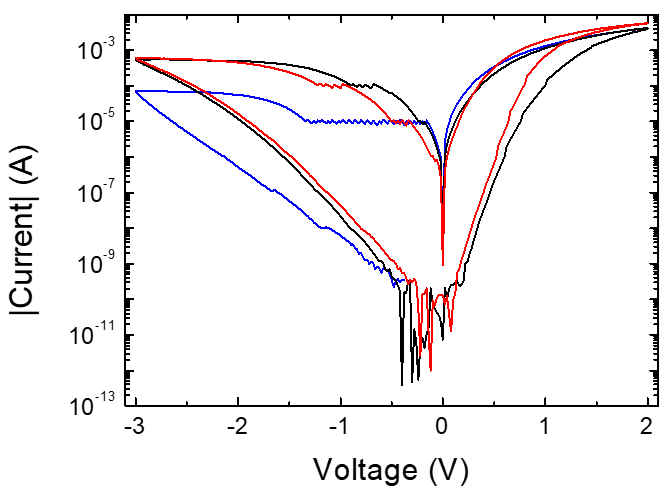}
	\caption{
	\textbf{Measurements of 800 nm devices: } device-to-device variation when controlled with a larger voltage range. The red graph is the device presented in the main text. These devices show a greater degree of variation, due to small differences in their areas and edges arising from the fabrication process. There resistance ratios, however remain high.
}
\label{fig:Figs4}
\end{figure}
\begin{figure}[H]
	\centering
	\includegraphics[width=\textwidth]{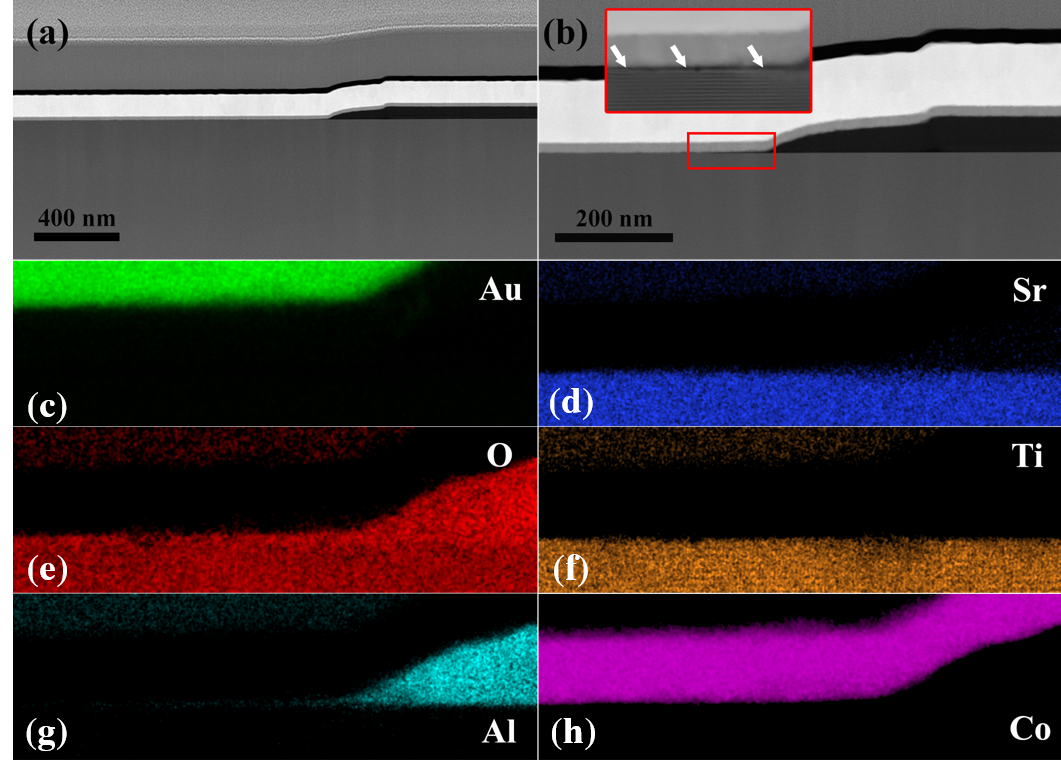}
	\caption{
\textbf{Side wall profile of electrical measurement device: }(a) and (b) STEM-HAADF images. The inset in (b) marks the interfacial region close to the edge. STEM-energy-dispersive X-ray spectroscopy (STEM-EDX) elemental mapping image of (c) Au, (d) Sr, (e) O, (f) Ti, (g) Al and (h) Co.
}
\label{fig:Figs5}
\end{figure}
\begin{figure}[H]
	\centering
	\includegraphics[width=\textwidth]{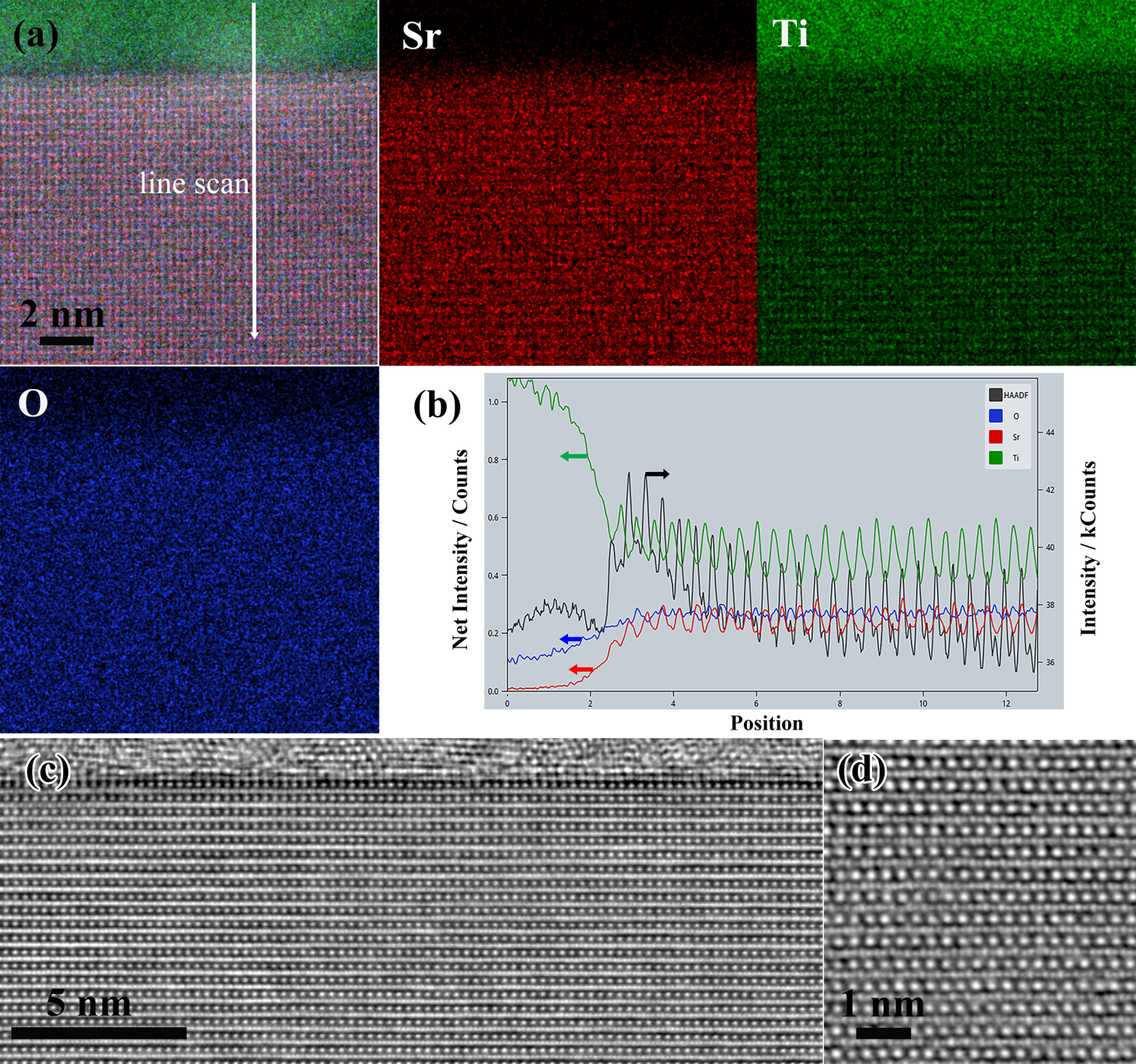}
	\caption{
\textbf{Nb:STO/Ti interface: }(a) STEM-EDX elemental map of Sr Ti and O. (b) elemental intensity as a function of position along the line scan in (a). STEM-iDPC images of (c) the interface and (d) away from the interface.
}
\label{fig:Figs6}
\end{figure}
\begin{figure}[H]
	\centering
	\includegraphics[width=0.9\textwidth]{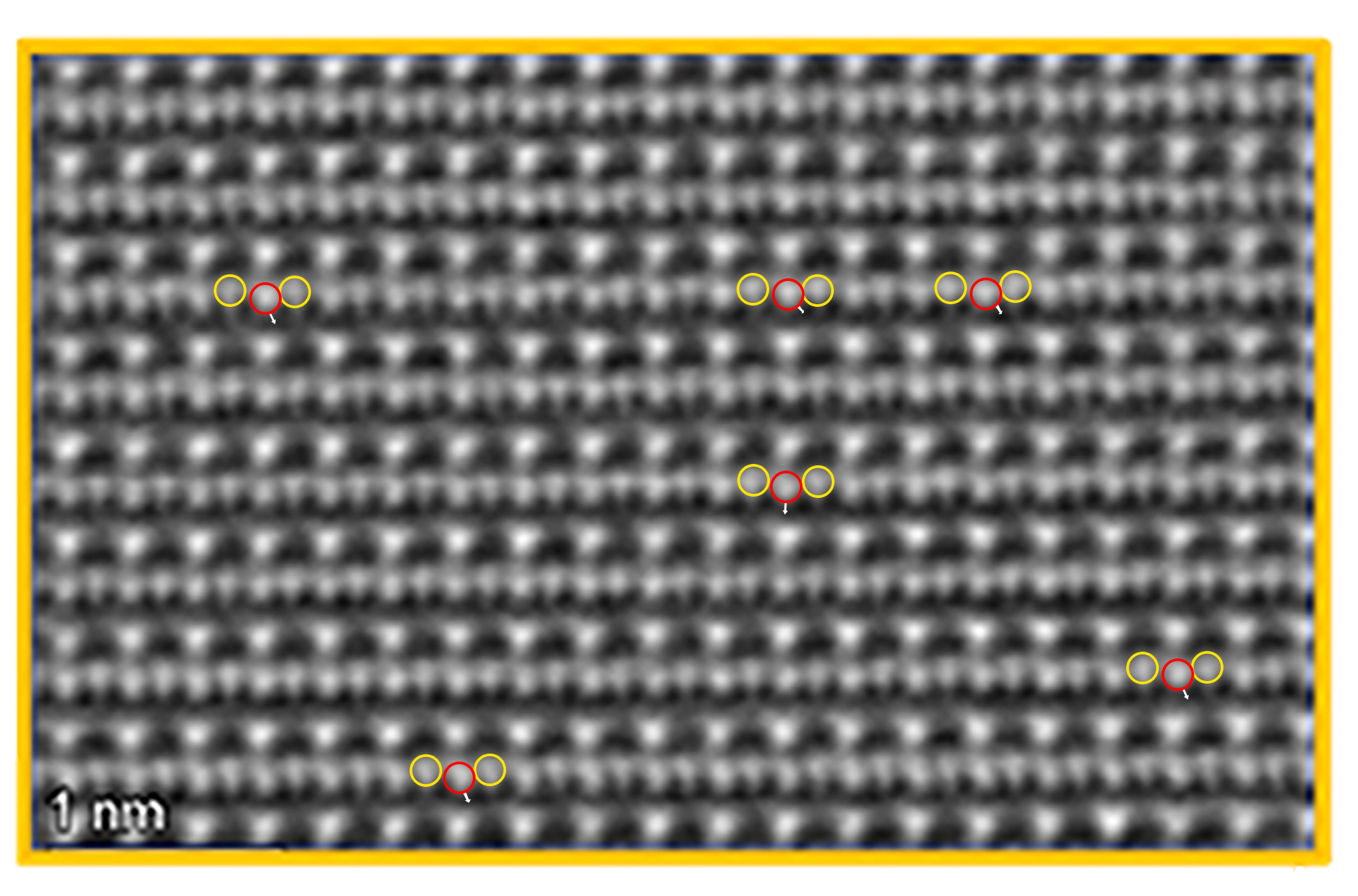}
	\caption{
\textbf{Ti-column displacement: }iDPC-STEM image inside Nb:STO substrate close to the interface. Some of the Ti ions occupying ideal perovskite positions are marked in yellow while displaced ions are marked in red with arrows highlighting the direction of displacement.
}
\label{fig:Figs7}
\end{figure}
\subsection*{S1. Derivation of the relation between trapping density and exponent.}
\label{S1}
\begin{figure}[H]
	\centering
	\includegraphics[width=\textwidth]{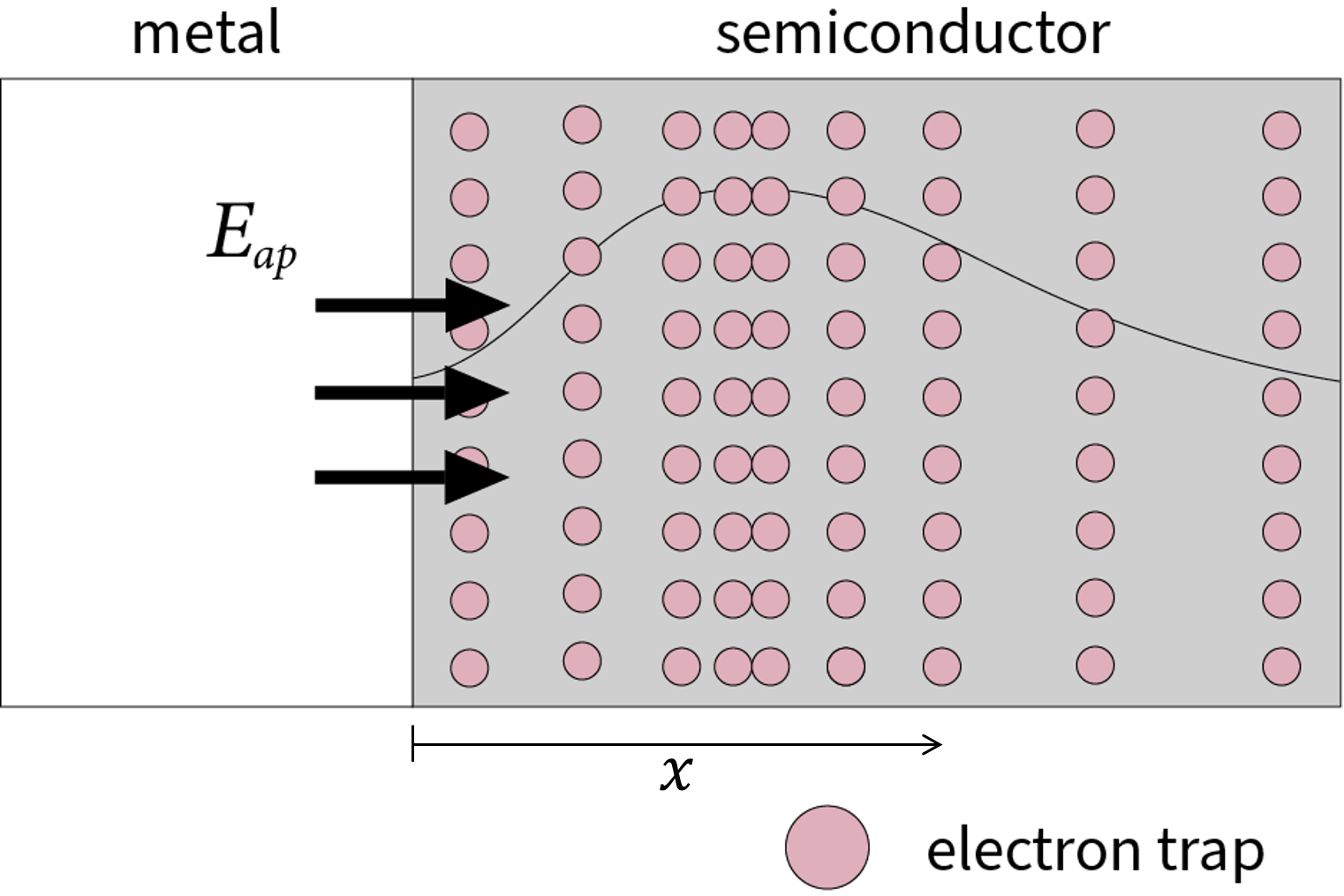}
	\caption{
\textbf{Schematic of parameters in section S1: }$E_{ap}$ and $x$ represent the applied electric field and centroid of trapped charge, defined with respect to the interface, respectively. The number density of trapped charges, $n$, is depicted by the black curve as a function of position in the dielectric.
}
\label{fig:Figs8}
\end{figure}

If we assume that the rate of trapping has no dependence on the location of traps, the electric field, $E$, can be expressed as:
\begin{equation}
\label{field}
    E=E_{ap}-\frac{qnx}{\eta}
\end{equation}
where $E_{ap}$ is the applied electric field, $q$ the electric charge, $n$ is the number density of trapped charges, $x$ is the centroid of the trapped charge with respect to the interface and $\eta$ is the dielectric permittivity. In \citeS{wolters1985}, charge trapping was analyzed on the basis of three mechanisms, namely first-order trapping, first-order trapping with Coulombic interactions, and trapping which increases during injection due to the generation of states. The expressions for current they derive are qualitatively similar for each mechanism. Hence, for simplicity, we consider the rate of trapping density to be a decay in first order with the addition of electron-electron interactions. Coulombic repulsion may inactivate trapping sites surrounding a trapped electron. This is included in the rate equation by multiplying a probability factor. If the volume of dielectric rendered inactive by a trap is $h$, then the trapping is reduced by a factor of $(1-\frac{h}{V})$, where $V$ is the volume of the dielectric. For $n$ trapped charges, the factor is $(1-\frac{h}{V})^n$. The trapping rate can be expressed as:
\begin{equation}
    \frac{dn}{dt}=(n_0-n)\sigma\frac{Jv_{th}}{qv_d}\Big(1-\frac{h}{V}\Big)^n
\end{equation}
where $n_0$ is the maximum number of traps available, $J/q$ is the net flux density, $v_{th}$ and $v_d$ are the thermal and drift velocities respectively, and $\sigma$ is capture cross-section. Assuming the total volume of the dielectric to be much larger than the volume deactivated by trapping events so that, 1$\gg h/V$ and $n_0\gg n$, this expression can be simplified to:
\begin{equation}
    \frac{dn}{dt}=n_0\sigma\frac{Jv_{th}}{qv_d}e^{-\frac{nh}{V}}
\end{equation}
Solving this equation yields the following expression for $n$:
\begin{equation}
\label{nSI}
    n=\frac{V}{h}\ln\Big(\frac{Q}{Q^*}+1\Big)
\end{equation}
where $Q=\int Jdt$ is the total injected charge and
\begin{equation}
    Q^*=\frac{Vv_dq}{n_0hv_{th}\sigma}
\end{equation}

We express the current in terms of the electric field as:
\begin{equation}
\label{lnJ}
    \ln\Big(\frac{J}{J_0}\Big)=\frac{E}{E_0}=\frac{1}{E_0}\Big(E_{ap}-\frac{V}{h}\frac{qx}{\eta}\ln\Big(\frac{Q}{Q^*}+1\Big)\Big)
\end{equation}
The current follows a decaying power law with time, $J=J_s t^{− \alpha}$, and the injected charge as
a function of time is given by:
\begin{equation}
\label{qt}
 Q(t)=\int Jdt= \frac{J_st^{1-\alpha}}{1-\alpha} 
\end{equation}
Substituting \ref{qt} into \ref{lnJ} when $Q/Q^*\gg$ yields and noting $\beta=\frac{Vqx}{hE_0\epsilon}$:
\begin{equation}
\label{comp}    
\begin{split}
         \ln\Big(\frac{J}{J_0}\Big)&=\frac{1}{E_0}\Big(E_{ap}-\beta\ln\Big(\frac{J_st^{1-\alpha}}{Q^*(1-\alpha)}\Big)\Big) \\
     &=\frac{1}{E_0}\Big(E_{ap}-\beta\ln\Big(\frac{J_st}{Q^*(1-\alpha)}\Big)-\beta(1-\alpha)\ln t\Big)
\end{split}
\end{equation}
Comparing \ref{comp} with $J=J_s t^{− \alpha}$ implies 
\begin{equation}
\label{alpha}
    \begin{split}
        \alpha &=\beta(1-\alpha) \\
        &=\frac{\beta}{1+\beta}
    \end{split}
\end{equation}
and 
\begin{equation}
    J_s\approx mE_{ap}+n-\beta\ln(J_s)
\end{equation}
Where $m$ encompasses several material parameters. Writing $\beta$ in terms of $\alpha$, and since measured currents are less than 10$^{−4}$ A, $J_s$ can be neglected in comparison
to $\ln(J_s)$, leading to:
\begin{equation}
    \frac{\alpha}{1-\alpha}\ln(J_s)\approx mE_{ap}+n
    \label{proof}
\end{equation}
$\beta$ is positive, we know from Eq. \ref{alpha} that $\alpha$ lies between 0 and 1, and is a monotonically increasing function of $\beta$. Considering that $\beta=\frac{V}{h}\frac{qx}{E_0}$, an increase in either the effective density, $V/h$ or in $x$ gives rise to an increase in $\alpha$, with the former being physically more likely. 

Instead of deriving an explicit expression for the number density of trapped charge, we can also directly relate the trapping rate to the current, as was done in for example \citeS{walden1972}. We use $Q_T$ to denote the charge that is trapped when charge $Q$ is injected into the dielectric. The ratio $\frac{dQ_T}{dQ}$ is assumed to be a function of current, i.e.
\begin{equation}
\label{QT}
    \frac{dQ_T}{dQ}=f\Big(\frac{J}{J_0}\Big)
\end{equation}
Substituting Eq. \ref{QT} into Eq. \ref{field} gives:
\begin{equation}
    \frac{dE}{dt}=\frac{Jx}{l\eta}\frac{dQ_T}{dQ}
\end{equation}
where $l$ is the length of the dielectric and $J=\frac{dQ}{dt}$. To relate this to the power law, we assume a solution of the form 
\begin{equation}
\label{power}
    \frac{dQ_T}{dQ}=\Big(\frac{J}{J_0}\Big)^{\alpha}
\end{equation}
with $\alpha\geq 0$. A general expression for the current assumes the form:
\begin{equation}
    J(E)=J_0(E)e^{g(E_0)}
\end{equation}
where the  specific functions are determined by the relevant conduction mechanisms. Specifically here, using Eq. \ref{power} we can express the current as
\begin{equation}
    J=J_s\frac{t}{t_0}^{-1/(\alpha +1) }
\end{equation}
For conduction given by an exponential relation as in Eq. \ref{lnJ}:
\begin{equation}
\label{expSI}
    J_s\propto e^{\frac{(1-1/(\alpha + 1)V}{V_0}}
\end{equation}
while for conduction determined by Frenkle-Poole equation, we arrive at:
\begin{equation}
\label{FP}
    J_s\propto Ve^{\frac{(1-1/(\alpha + 1)V^{\frac{1}{2}}}{V_0}}
\end{equation}
and for Fowler-Nordheim conduction we get:
\begin{equation}
\label{FN}
    J_s\propto V^2e^{\frac{(1-1/(\alpha + 1))V^{-1}}{V_0}}
\end{equation}
Here, $V_0$ is a constant. 

The dominant transport mechanism in a system can be determined by plotting the current versus voltage on a double logarithmic scale. Equations \ref{alpha}, \ref{proof}, \ref{expSI}, \ref{FP} and \ref{FN} indicate a clear theoretical proof that the exponent in the power law is directly proportional to the effective trap density or capacity of the dielectric to trap electrons, independent of which transport mechanism is dominant.
\newpage
\begin{figure}[H]
	\centering
	\includegraphics[width=\textwidth]{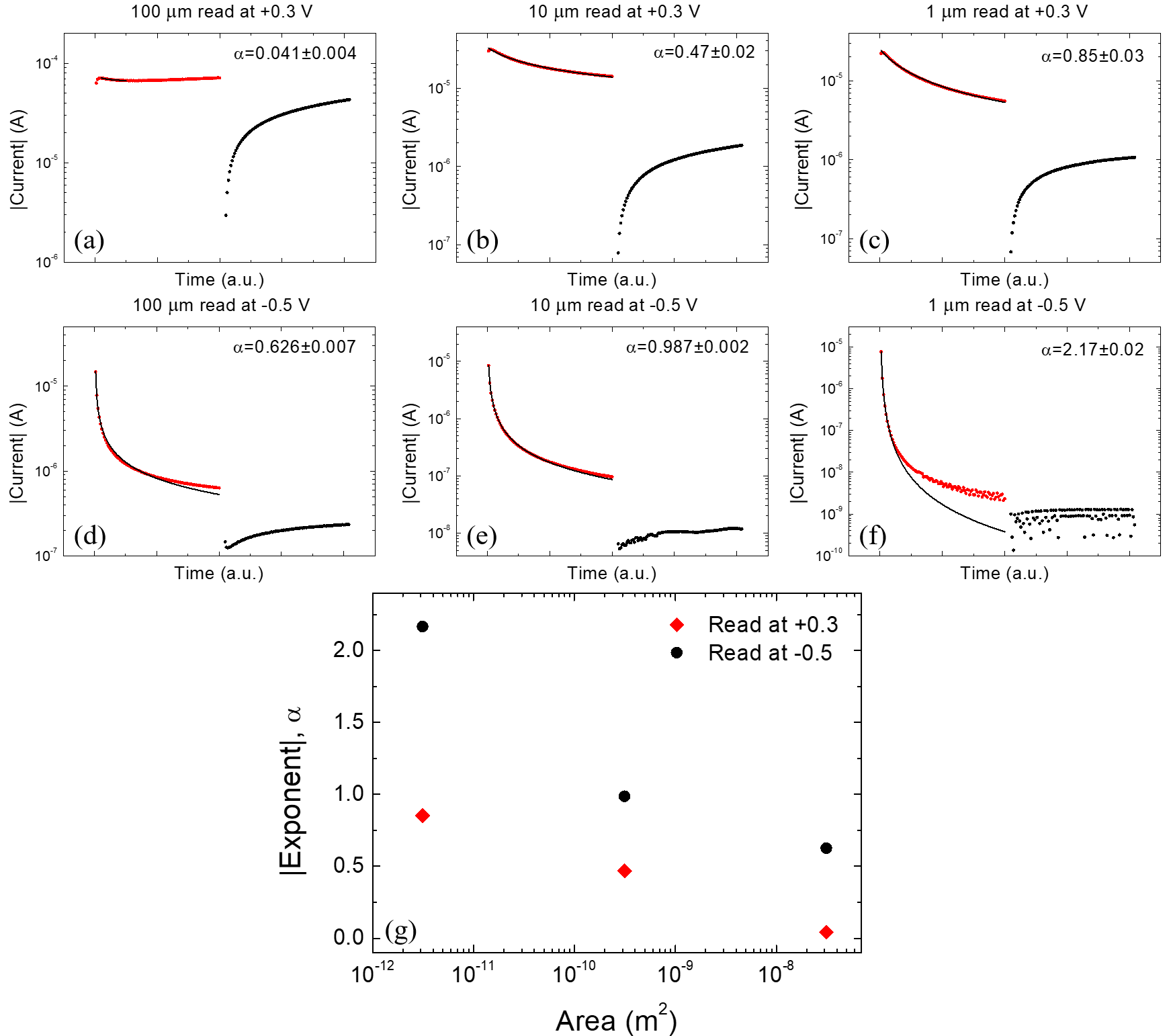}
	\caption{
\textbf{Fits of the retention data to extract the exponents $\alpha$: }The model used is $|I|=I_0(t-t_0)^{-\alpha})$, where $|I|$ and $t$ are the absolute current and time respectively, and $I_0$ and $t_0$ are fitting parameters. The adjusted $R^2$ values of the fits are (a) 0.99237, (b) 0.99475, (c) 0.99689, (d) 0.99717, (e) 0.99995, and (f) 0.99996. (g) shows the dependence of the exponents on area.
}
\label{fig:Figs9}
\end{figure}

\subsection*{S2. Modeling the edge effects}
\label{S2}
To visualize the field profiles in our devices we used finite element analysis (COMSOL). The modeling geometry is shown in Fig S10. In each simulation, the Nb:STO substrate was modelled as a cube with a dielectric constant of 300 and a thickness of 0.5 mm (along z), corresponding to the thickness used in the experimental study. A circular Co electrode of radius 1 \textmugreek m, 10 \textmugreek m or 100 \textmugreek m was placed on the top surface of the substrate (z=0.5 mm). A ground node was placed on the bottom of the substrate (z=0), while a voltage was applied to the top Co electrode. For the simulations in Fig. \ref{fig:Figs13}. the size of the substrate was reduced to improve the resolution of the mesh. This was required to retain the circular nature of the electrodes for the 10 nm devices; this was determined not to influence the electric field strength. 
\begin{figure}[H]
	\centering
	\includegraphics[width=0.9\textwidth]{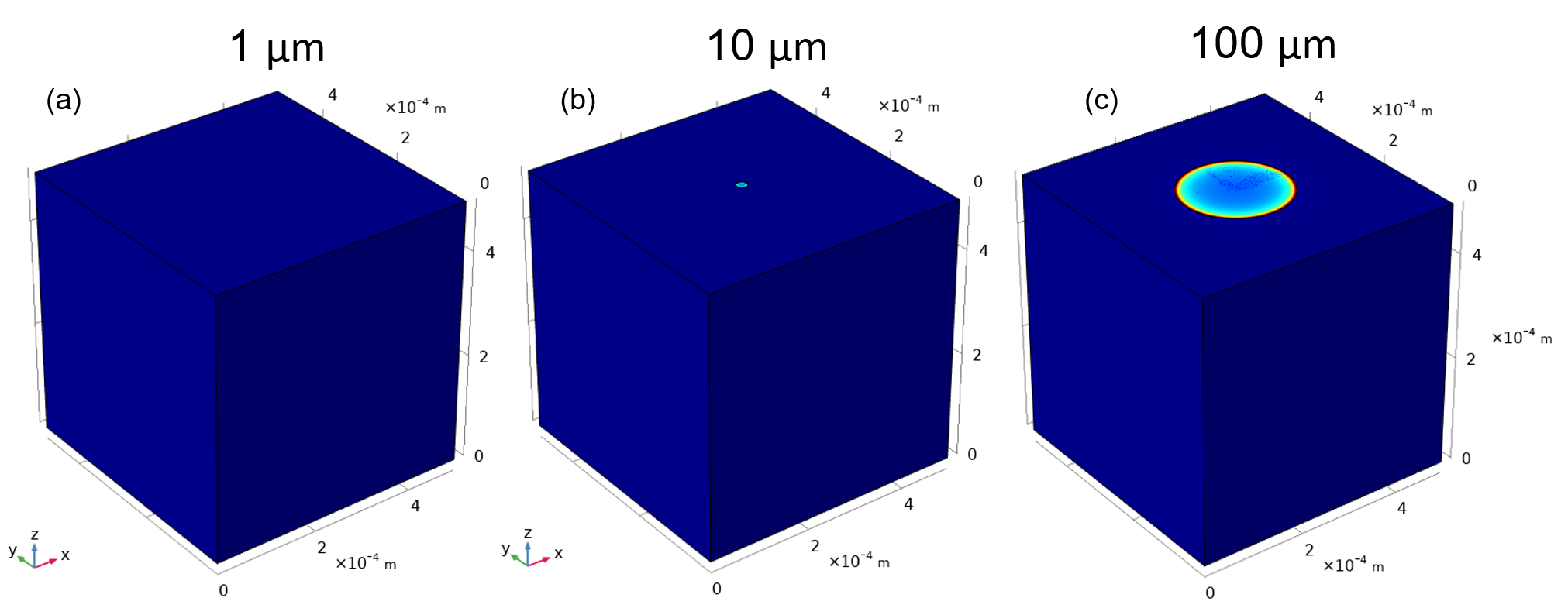}
	\caption{
	\textbf{Model sample geometry:} the Nb:STO substrate is represented by a cube with a thickness of 0.5 mm (along z). Circular Co electrode of radii (a) 1 \textmugreek m, (b) 10 \textmugreek m and (c) 100 \textmugreek m is placed on the top surface of the substrate (z=0.5 mm). A ground node is placed on the bottom of the substrate (z=0), while a voltage is applied to the top Co electrode.
}
\label{fig:Figs10}
\end{figure}

\begin{figure}[H]
	\centering
	\includegraphics[width=\textwidth]{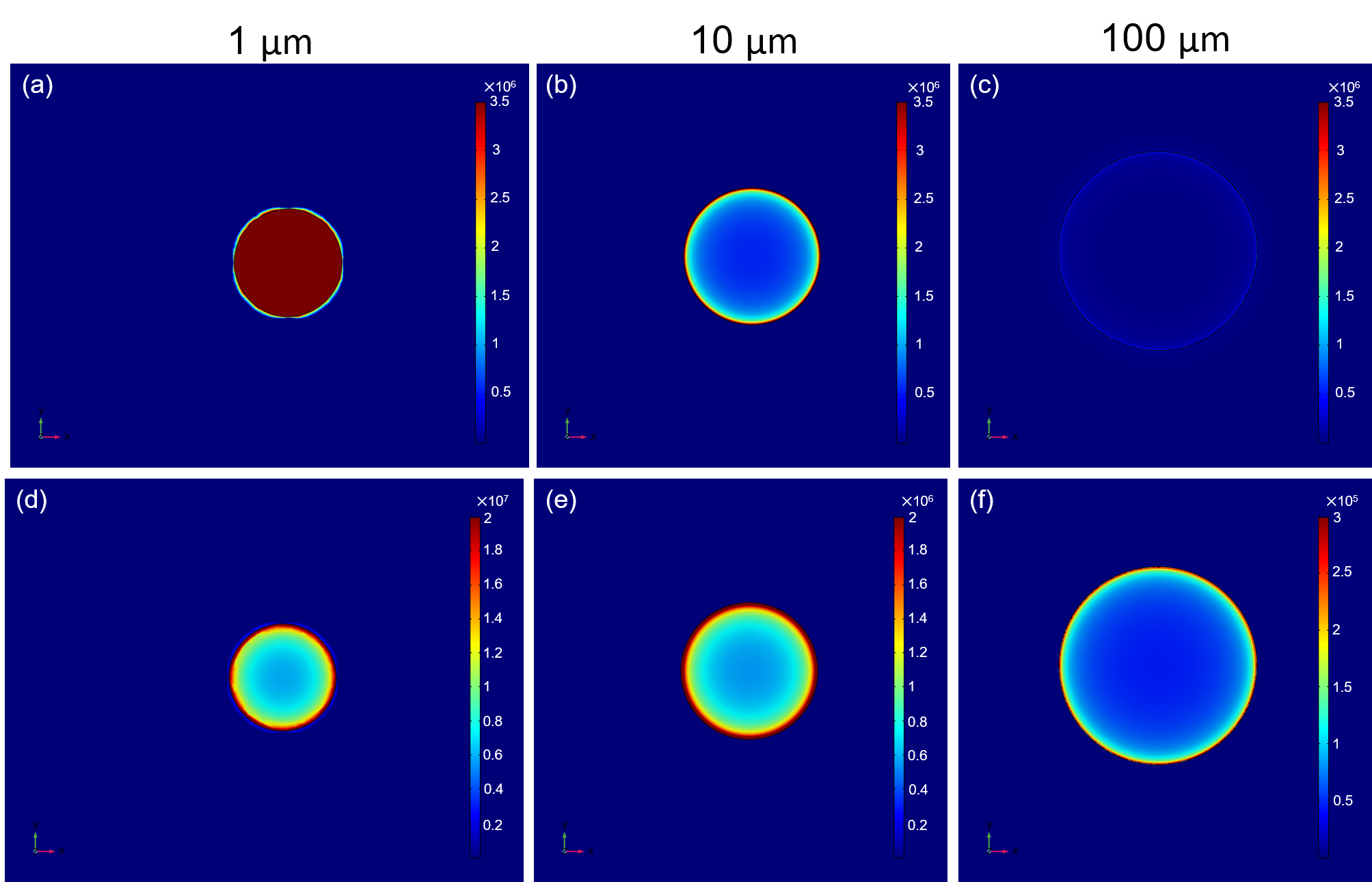}
	\caption{
	\textbf{Electric field at -3 V: } along the surface normal (z direction) for (a)+(d) 1 \textmugreek m, (b)+(e) 10 \textmugreek m and (c)+(f) 100 \textmugreek m devices, The plots on the top row ((a)-(c)) have the same scale bar, with a maximum field value of 3.5 $\times$10$^6$ Vm$^{-1}$. For the bottom row, the scale bar has a maximum value of (d) 2 $\times$10$^7$ Vm$^{-1}$, (e) 2 $\times$10$^6$ Vm$^{-1}$ and (f) 3 $\times$10$^5$ Vm$^{-1}$.
}
\label{fig:Figs11}
\end{figure}
\begin{figure}[H]
	\centering
	\includegraphics[width=\textwidth]{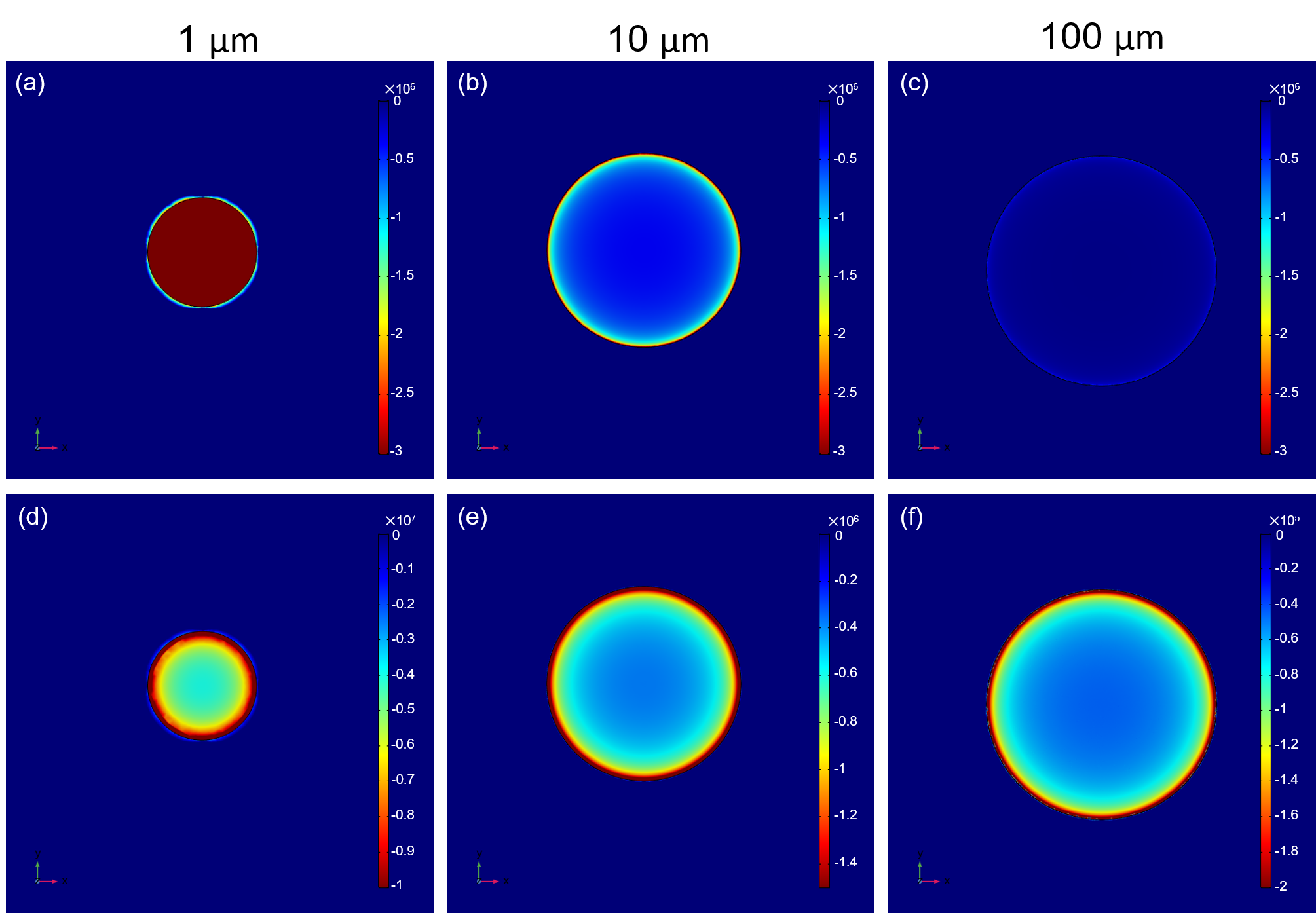}
	\caption{
	\textbf{Electric field at +2 V: } along the surface normal (z direction) for (a)+(d) 1 \textmugreek m, (b)+(e) 10 \textmugreek m and (c)+(f) 100 \textmugreek m devices, The plots on the top row ((a)-(c)) have the same scale bar, with a maximum field strength of -3 $\times$10$^6$ Vm$^{-1}$. For the bottom row, the scale bar has a maximum value of (d) -1 $\times$10$^7$ Vm$^{-1}$, (e) -1.5 $\times$10$^6$ Vm$^{-1}$ and (f) -2 $\times$10$^5$ Vm$^{-1}$.
}
\label{fig:Figs12}
\end{figure}
\begin{figure}[H]
	\centering
	\includegraphics[width=\textwidth]{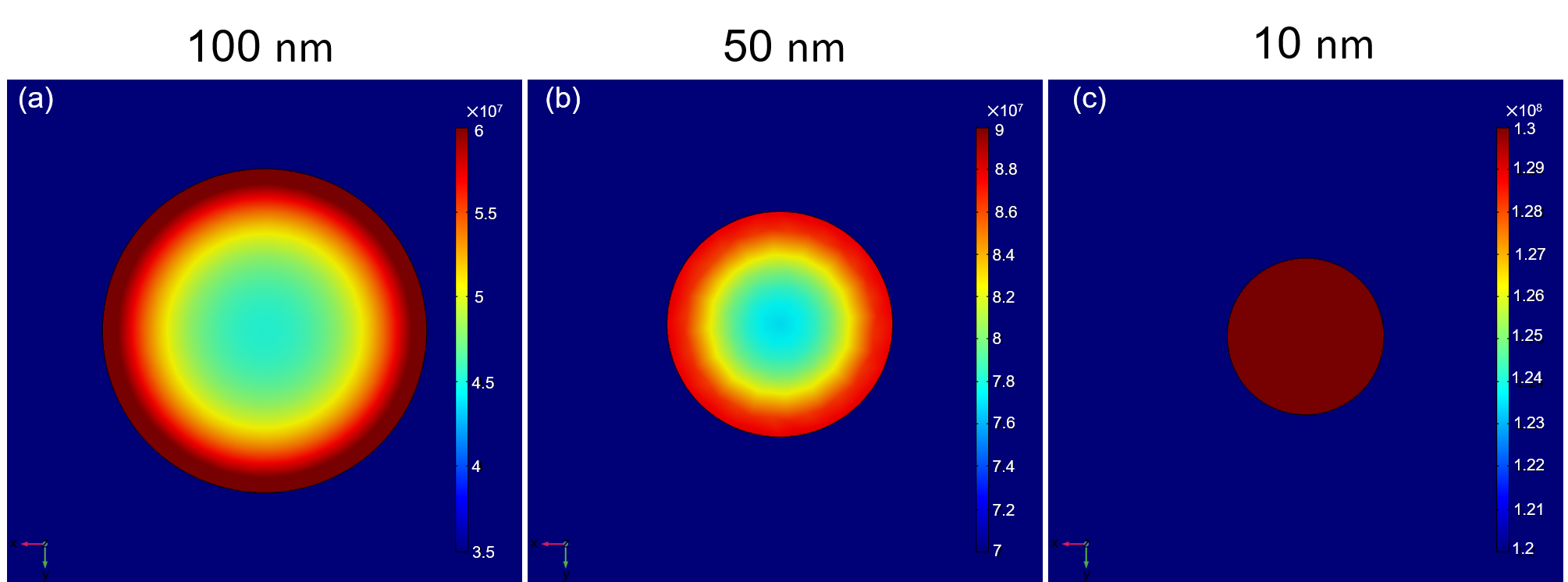}
	\caption{
	\textbf{Electric field at -3 V: } along the surface normal (z direction) for (a) 100 nm, (b) 50 nm and (c) 10 nm devices. No saturation of the field is observed in (a) and (b) and the field appears to saturate in the 10 nm devices.
}
\label{fig:Figs13}
\end{figure}
\newpage
\subsection*{S3. Literature survey of interfacial switching}
\label{S3}
It is often suggested that a layer close to the interface layer is responsible for the switching \citeS{mikheev2014,bourim2014,fan2017}. Some groups have shown that both high and low resistance states show an area-independent current density, eluding to a switching mechanism that occurs homogeneously over the entire device area \citeS{sim2005}. Often this is explained in terms of a change in the Schottky barrier height and width induced by charge trapping at the interface \citeS{ni2007,mikheev2014,bourim2014,yin2015,goossens2018,li2018} and movement of oxygen vacancies \citeS{yin2015,li2018}. Other explanations are proposed where the barrier profile is unchanged and interfacial changes happen at local regions. Explanation of this type includes
It has also been proposed that the application of a positive bias results in the generation of oxygen vacancies, forming tunnelling paths and giving rise to a LRS where tunnelling, rather than thermionic emission dominate charge transport. The application of a negative bias results in the accumulated of large amounts of oxygen in the vacancies which prevents tunnelling and gives a HRS \citeS{fujii2007,seong2008}. 

Rodenb{\"u}cher \textit{et al.} used local-conductivity AFM on highly doped Nb:STO to show the presence of nanoscale conducting and switchable clusters. Suggesting that in this case switching is a local phenomenon related to the presence of conducting clusters with higher Nb content than their surroundings \citeS{rodenbucher2013}.

Finally Chen \textit{et al.} used scanning tunnelling microscopy and spectroscopy to study the resistive switching in Nb-doped SrTiO$_3$ without an electrode, demonstrating that oxygen migration is the results in a variation of electronic structure during the switching. With a negative voltage, oxygen anions at the interface near the STM tip were oxidised into oxygen molecules and left the lattice. Simultaneously, oxygen vacancies diffuse into the sample, which act like donor-like levels causing distortions in LDOS near conduction band and enhance the carrier concentration with electron hopping, thus increasing the sample’s conducting. With a positive voltage, oxygen anions return into the sample and the influence of the donor-like level became weak and the conductivity decreased \citeS{chen2012}.

Despite a large number of contradictory results and explanations, factors of importance that have been identified include the semiconductor doping concentration, electrode material and the quality of the interface.
\newpage

\bibliographystyleS{unsrt}
\bibliographyS{arxiv}

\end{document}